\documentclass[%
 reprint,
 showpacs,preprintnumbers,
 amsmath,amssymb,amsfonts,
 aps,
prb
]{revtex4-1}
\usepackage{dcolumn}
\usepackage[dvipdfmx]{graphicx}
\usepackage{subfigure}
\usepackage[dvipdfmx]{color}
\usepackage{mathrsfs}
\usepackage{bbm}

\makeatletter
\def\btt#1{\texttt{\@backslashchar#1}}
\DeclareRobustCommand\bblash{\btt{\@backslashchar}} \makeatother

\def\gsim{\lower -0.3ex \hbox{$>$} \kern -0.75em \lower 0.7ex
	\hbox{$\sim$}}
\def\lsim{\lower -0.3ex \hbox{$<$} \kern -0.75em \lower 0.7ex
	\hbox{$\sim$}}

\def\Vec#1{\textbf{#1}}
\newcommand{\GVec}[1]{\mbox{\boldmath$#1$}}

\begin{document}

 \title{
Magnetic susceptibility in three-dimensional nodal semimetals
 }

 \author{Mikito Koshino and Intan Fatimah Hizbullah}
 \affiliation{Department of Physics, Tohoku University, Sendai 980-8578, Japan}

 \date{\today}

 \begin{abstract}
We study the magnetic susceptibility in various three-dimensional
gapless systems, including Dirac and Weyl semimetals and a line-node semimetal.
The susceptibility is decomposed into 
the orbital term, the spin term and also the spin-orbit cross term which is caused by the spin-orbit interaction.
We show that the orbital susceptibility logarithmically diverges
at the band touching energy in the point-node case,  
while it exhibits a stronger delta-function singularity in the line node case.
The spin-orbit cross term is shown to be paramagnetic in the electron side
while diamagnetic in the hole side,
in contrast with other two terms which are both even functions in Fermi energy.
The spin-orbit cross term in the nodal semimetal
is found to be directly related to the chiral surface current induced by the topological surface modes.
 \end{abstract}

 \pacs{75.20.-g, 73.20.At, 03.65.Vf}


 \maketitle

\section{Introduction}
\label{chap:intro}

Recently, a great deal of attention has been focused on the three-dimensional (3D) materials having 
zero-gap electronic structure with nontrivial topology.
The Dirac semimetal is a representative example of such gapless materials,
where four energy bands are touching at an isolated point in the momentum space,
under the protection of the time-reversal and the spatial inversion symmetry.
\cite{murakami2007phase, young2012dirac,wang2012dirac}
By breaking either symmetry, the four-fold degeneracy splits into a pair of doubly degenerate Weyl-nodes
and then the system is called the Weyl semimetal.
\cite{murakami2007phase, burkov2011weyl, burkov2011topological, wan2011topological, hosur2013recent}
The realizations of the Dirac and Weyl semimetal phases 
have been reported in the recent experiments.
\cite{liu2014discovery,liu2014stable,brahlek2012topological,borisenko2014experimental,xu2015observation,
weng2015weyl,lv2015experimental,huang2015weyl,xu2015discovery}
Theoretically, it is also possible to have the band touching on a line in the momentum space,
and such line-node band models and candidate systems were proposed. \cite{burkov2011topological,phillips2014tunable,kim2015dirac,yu2015topological,weng2015topological,ramamurthy2015quasi}
The response of the nodal semimetals to the electromagnetic field attracts a great interest, and in particular, 
the exotic magnetoelectric effect related to its topological nature 
has been a topic of extensive research.  \cite{aji2012adler,son2012berry,zyuzin2012topological,vazifeh2013electromagnetic,goswami2013axionic,hosur2013recent,son2013chiral,liu2013chiral,burkov2014chiral}

In this paper, we study the magnetic susceptibility of the three-dimensional nodal semimetals
and show that it also exhibits various unusual properties.
In two-dimenisional (2D) zero-gap linear band (e.g., graphene), it is known that the orbital susceptibility is expressed 
as a delta function of the Fermi energy, which diverges in the negative (i.e.,  diamagnetic) direction
at the band touching point. 
\cite{mcclure1956diamagnetism,safran1979theory,koshino2007diamagnetism,koshino2007orbital}
A similar calculation was also done for 3D Dirac Hamiltonian,
and there the susceptibility exhibits a weaker logarithmic singularity 
in the gapless limit.  \cite{koshino2010anomalous}
Here we calculate the magnetic susceptibility in a wider variety of three-dimensional
gapless systems, including Dirac and Weyl semimetals as well as the line-node semimetal.
The calculation is not a simple extension of the previous works 
in that we deal with the contribution from the spin degree of freedom.
The 3D nodal band structure often originates from a strong spin-orbit interaction,
and in such a situation the magnetic susceptibility contains the spin-orbit cross term 
in addition to the usual orbital susceptibility and the spin susceptibility. \cite{ItoNomura}
Here we consider a parameterized $4\times 4$ continuum Hamiltonian which covers 
the point node and line node semimetals as well as conventional gapped semiconductors,
and calculate these susceptibility components.

We show that the orbital susceptibility logarithmically diverges
at the band touching energy in the point-node semimetals,  \cite{koshino2010anomalous}
while it exhibits a stronger delta-function singularity in the line-node case.
The spin susceptibility, which describes the Pauli and Van-Vleck paramagnetism,
is found to be completely Fermi-energy independent 
near the degenerate points in the point-node case.
The spin-orbit cross term is shown to be an anti-symmetric function of the Fermi energy,
which is paramagnetic in the electron side and diamagnetic in the hole side.
In the nodal semimetals, the spin-orbit cross term is found to be directly related to 
the orbital magnetization induced by the topological surface modes.
Besides the continuum model, we also consider the corresponding lattice model to
confirm the validity of the continuum approximation.

The paper will be organized as follows.
We introduce the continuum model and the lattice model for the point-node / line-node semimetals
in Sec.\ \ref{sec_model}.
The calculation of the magnetic susceptibility is presented 
in Sec.\ \ref{sec_mag}
where the point-node case and the line-node case
are separately argued in Secs.\ \ref{sec_mag}A and \ref{sec_mag}B,
respectively.
A brief conclusion is given in Sec.\ \ref{sec_concl}.

\begin{figure*}
\begin{center}
 \leavevmode \includegraphics[width=0.8\hsize]{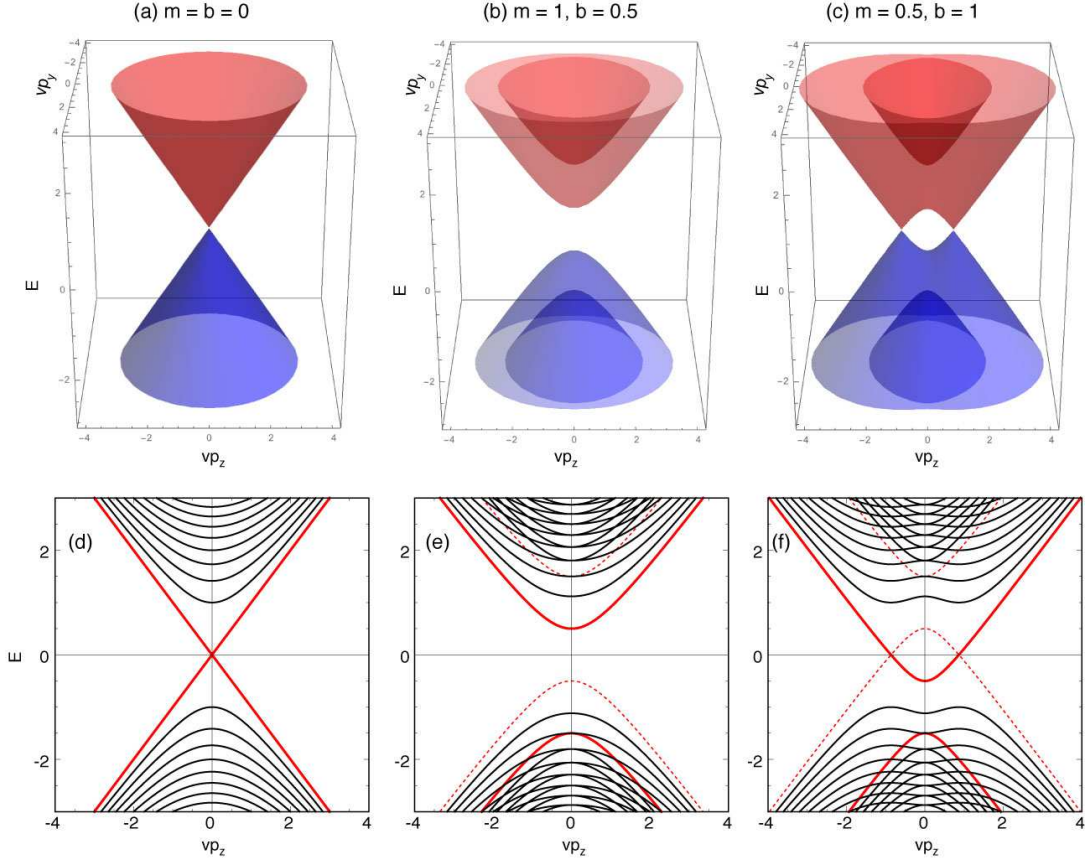}
        \caption{
        (Above) Energy spectra of $\varepsilon_{s\mu}(0,p_y,p_z)$ 
        for Dirac semimetal case ($m=b=0$), semiconducting case ($m=1, b=0.5$)
        and Weyl semimetal case ($m=0.5, b=1$) for the Hamiltonian Eq.\ (\ref{H}).
        (Below)   Corresponding Landau level spectra at the magnetic field $\Delta_B=1$. 
        The thick red curve represents the zero-th Landau level,
        while the dashed curve non-existing zero-th level  (see the text). 
        }
\label{fig_spec}
\end{center}
\end{figure*}

\section{Model Hamiltonian and energy spectrum}
\label{sec_model}

\subsection{Point-node semimetal}
\label{sec_point}

We consider a $4\times 4$ Hamiltonian matrix,
\begin{align}
H &= v\tau_x (\GVec{\sigma}\cdot\Vec{p}) + m\tau_z + b\sigma_z
\nonumber\\
&=
\begin{pmatrix}
   m \mathbbm{1} + b\sigma_z & v \GVec{\sigma}\cdot\Vec{p} \\
   v \GVec{\sigma}\cdot\Vec{p} & -m \mathbbm{1} + b\sigma_z
 \end{pmatrix},
\label{H}
\end{align}
where $\Vec{p}=(p_x,p_y,p_z)$ is the momentum vector,
$\GVec{\sigma}=(\sigma_x,\sigma_y,\sigma_z)$ is the Pauli matrices
for the spin degree of freedom,
$(\tau_x,\tau_y,\tau_z)$ is the Pauli matrices for pseudospins
corresponding to, e.g., sublattices or atomic orbitals,
and $\mathbbm{1}$ is the $2\times 2$ unit matrix.
The model includes the velocity parameter $v$,  the mass parameter $m$,
and the intrinsic Zeeman field $b$ which can arise in a magnetic material
breaking time-reversal symmetry.
By diagonalizing the Hamiltonian, we obtain four energy bands,
\begin{equation}
\varepsilon_{s\mu}(\Vec{p}) = s \sqrt{m^2+b^2+v^2 p^2 + \mu  2b\sqrt{v^2 p_z^2 + m^2}},
\end{equation}
where $p=|\Vec{p}|$, 
$s=\pm$ and $\mu=\pm$.
The energy spectrum $\varepsilon_{s\mu}(0,p_y,p_z)$ is plotted 
in Fig.\ \ref{fig_spec} for (a) $m=b=0$, (b) $|m|>|b|$ and (c) $|b|>|m|$.
The case of $m=b=0$ corresponds to the Dirac semimetal, where the spectrum is
composed of a pair of degenerate linear bands touching at $\Vec{p}=0$.
The case  $|m|>|b|$ describes the semiconducting spectrum
where the energy band is gapped in the range  $|E| < |m|-|b|$.
Finally, $|b|>|m|$ represents the Weyl semimetal where the the middle two bands 
touch at a pair of isolated point-nodes $\Vec{p} = (0,0,\pm \sqrt{b^2-m^2}/v)$. 


In the presence of the external magnetic field $B$ in $z$ direction, 
the Hamiltonian (\ref{H}) becomes,
\begin{equation}
H = v\tau_x (\GVec{\sigma}\cdot\GVec{\pi}) + m\tau_z + \left(b + \frac{1}{2}g\mu_B B\right)\sigma_z  
\label{H_B}
\end{equation}
where $\GVec{\pi}=\mathbf{p}+(e/c)\mathbf{A}$,
$\mathbf{A}$ is vector potential to give $\mathbf{B}=\nabla\times \mathbf{A}$,
$g$ is  $g$-factor and $\mu_B = e\hbar/(2m_0c)$ is the Bohr magneton, and $m_0$ is the bare electron mass.
Here the external magnetic field $B$ enters the Hamiltonian 
in two different ways, one in the orbital part $\mathbf{p}+(e/c)\mathbf{A}$,
and the other in the spin Zeeman term $g\mu_B B \sigma_z$.
In the original Dirac equation for a relativistic electron, 
the magnetic field appears only through $\mathbf{A}$,
and the spin Zeeman term $g\mu_B B$ emerges out of it when
being reduced to the low-energy quadratic Hamiltonian. 
In the present Dirac-like equation in the solid state material, in contrast,
we need to add $g\mu_B B$ separately from $\mathbf{A}$.

The Landau levels can be found by using raising and lowering operators,
$\pi_x+i\pi_y = (\sqrt{2}\hbar/l_B)a^\dagger$
and $\pi_x-i\pi_y = (\sqrt{2}\hbar/l_B) a$,
which operate on the usual Landau-level wave function
$\phi_n$ in such a way that $a \phi_n = \sqrt{n}\phi_{n-1}$ and
$a^\dagger \phi_n = \sqrt{n+1}\phi_{n+1}$.
Here $l_B=\sqrt{c\hbar/(eB)}$ is the magnetic length.
For $n\ge 1$, the eigenfunction can be written as
\[
c_1 \phi_{n-1}\lvert\uparrow,\uparrow\rangle+  
c_2 \phi_{n} \lvert\downarrow,\uparrow \rangle+
c_3 \phi_{n-1}\lvert\uparrow,\downarrow \rangle+
c_4 \phi_{n}\lvert\downarrow,\downarrow \rangle,
\]
where $|s,s'\rangle$ represents 
$|\sigma_z=s\rangle\otimes| \tau_z=s'\rangle$.
The Hamiltonian matrix for the vector $(c_1,c_2,c_3,c_4)$ then becomes
\begin{equation}
H_{n\geq 1}= 
\begin{pmatrix}
   m+b& 0 & v p_z&\Delta_B\sqrt{n}\\
   0 & m-b &\Delta_B\sqrt{n}&-v p_z\\ v p_z&\Delta_B\sqrt{n}&-m+b&0\\\Delta_B\sqrt{n}&-v p_z&0&-m-b
  \end{pmatrix},
\end{equation}
where
\begin{equation}
\Delta_B = \sqrt{2}\hbar v/l_B,
\end{equation}
and $b$ is actually $b + g\mu_BB$. 
This gives the four energy levels
\begin{align}
\varepsilon_{s\mu p_z n} = s \sqrt{m^2+b^2+v^2 p_z^2+\Delta_B^2 n+\mu 2b\sqrt{v^2 p_z^2 + m^2}}&
\label{eq:landau}
\end{align}
where $s=\pm$, $\mu=\pm$ and $n = 1,2,3,\cdots$.
The $n=0$ is a special case where the eigenfunction is written as
$c_2 \phi_{n} \lvert\downarrow,\uparrow \rangle+
c_4 \phi_{n}\lvert\downarrow,\downarrow \rangle$.
The Hamiltonian matrix for  $(c_2,c_4)$ becomes
\begin{equation}
H_{n=0}=\begin{pmatrix}
   m-b & v p_z\\ v p_z&-m-b
   \end{pmatrix},
\end{equation}
which gives the only two energy levels
\begin{equation}
\varepsilon_{\mu p_z 0} = -b - \mu \sqrt{v^2 p_z^2 + m^2},
\label{eq:landau_0}
\end{equation}
with $\mu=\pm$ and no $s$ index.

The Landau level spectrum is plotted against $p_z$  in Figs.\ \ref{fig_spec} (d), (e) and (f)
for $m=b=0$, $|m|>|b|$ and $|b|>|m|$, respectively.
We note that the Landau levels of $n\geq 1$ [Eq.\  (\ref{eq:landau})]  are electron-hole symmetric
with respect to the zero energy because of the coexistence of $s=\pm$ branches.
The $n = 0$ level $\varepsilon_{\mu p_z 0}$ [Eq.\  (\ref{eq:landau_0}); indicated by the thick red curve] 
is not symmetric except for $b =  0$,
because the inverted level $-\varepsilon_{\mu p_z 0}$ does not actually exist  (dashed curve).
A set of the existing and non-existing levels, $(\varepsilon_{\mu p_z 0},-\varepsilon_{\mu p_z 0})$,
can be regarded as
 the $n=0$ sector of Eq.\ (\ref{eq:landau}).
More precisely, Eq.\ (\ref{eq:landau})
can be simplified  to  $s \lvert b + \mu \sqrt{v^2 p_z^2 + m^2}\rvert$ at $n=0$,
and it is equal to either of $\varepsilon_{\mu p_z 0}$ or $-\varepsilon_{\mu p_z 0}$.
We define the index $\tau_{s\mu p_z}=\pm 1$ by
\begin{eqnarray}
\varepsilon_{s\mu p_z 0} = \tau_{s\mu p_z}\varepsilon_{\mu p_z 0}. 
\label{eq_zeroLL}
\end{eqnarray}

Therefore, the complete set of the Landau level spectrum including zero-th level
is obtained by a single expression of Eq.\ (\ref{eq:landau}),
where the index $n$ of the sector $(s,\mu,p_z)$
runs from 0(1) when  $\tau_{s\mu p_z} = +1(-1)$.
This formulation will be used in deriving the magnetic susceptibility in the following section.

\begin{figure}
\begin{center}
 \leavevmode \includegraphics[width=0.65\hsize]{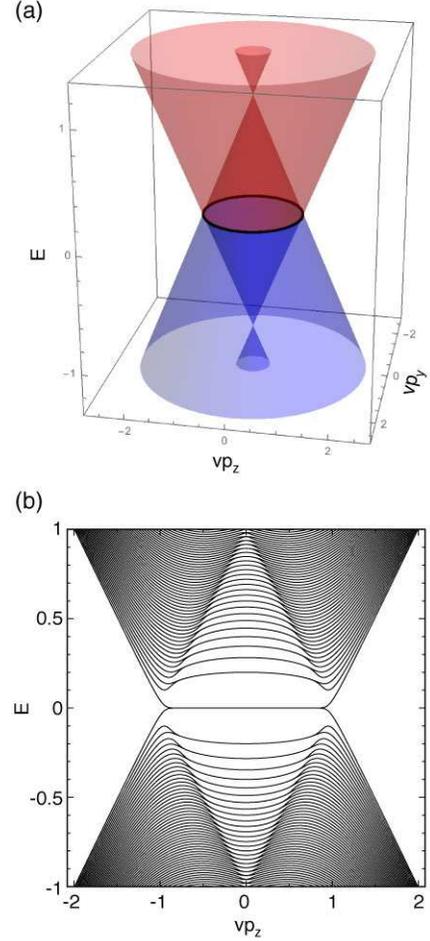}
        \caption{
         (Above) Energy spectrum of $\varepsilon_{s\mu}(0,p_y,p_z)$ 
         for the line-node semimetal described by Eq.\ (\ref{H_line}) with $b'=1$. 
        (Below)   Corresponding Landau level spectrum at the magnetic field $\Delta_B=0.2$. 
                }
\label{fig_spec_line}
\end{center}
\end{figure}

\begin{figure}
\begin{center}
 \leavevmode \includegraphics[width=0.7\hsize]{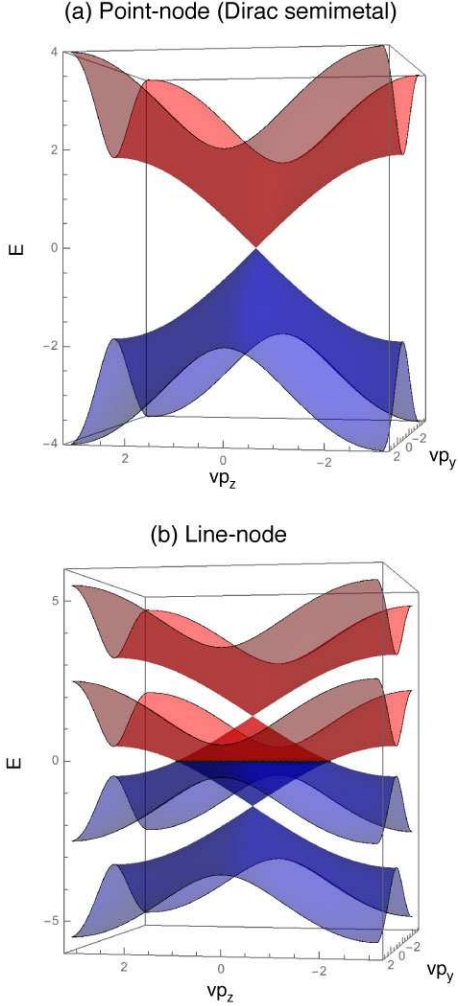}
        \caption{
        Energy spectra of $\varepsilon_{s\mu}(0,p_y,p_z)$ of the lattice models
         for (a) the Dirac semimetal [Eq.\ (\ref{H_lattice}) with $m=b=0$ and $u=r=1$] 
         and (b) the line-node semimetal [Eq.\ (\ref{H_line_lattice}) with  $b'=1.5$ and $u=r=1$] .
        }
\label{fig_spec_lattice}
\end{center}
\end{figure}

\subsection{Line-node semimetal}
\label{sec_line}

A line-node semimetal can be 
described by a similar Hamiltonian but with a different constant matrix.
Here we consider a particular model described by \cite{burkov2011topological}, 
\begin{align}
H &= v\tau_x(\GVec{\sigma}\cdot\Vec{p})+ b' \tau_z\sigma_x
=\begin{pmatrix}
    b'\sigma_x  & v \GVec{\sigma}\cdot\Vec{p}  \\
   v \GVec{\sigma}\cdot\Vec{p} & -b'\sigma_x  
 \end{pmatrix},
\label{H_line}
\end{align}
The last term $b' \tau_z\sigma_x$ represents the spin Zeeman field parallel to $x$,
of which direction is opposite between $\tau_z=+$ sector and $\tau_z=-$ sector.
The energy bands are given by
\begin{equation}
\varepsilon_{s\mu}(\Vec{p}) = s \sqrt{v^2 p_x^2 + \left[v\sqrt{p_y^2+p_z^2} + \mu b' \right]^2},
\label{E_line}
\end{equation}
with indexes $s=\pm$ and $\mu=\pm$
where the zero-energy contour
becomes a circle given by $p_x=0$ and  $\sqrt{p_y^2+p_z^2}=b' /v$.
Figure \ref{fig_spec_line}(a) shows the band structure at $p_x=0$
plotted against $(p_y,p_z)$, where the line node is indicated by a solid circle.
The spectrum is immediately gapped when $p_x$ goes away from 0.

The Hamiltonian under the external magnetic field $B$ in $z$ axis becomes
\begin{equation}
H = v(\GVec{\sigma}\cdot\GVec{\pi})\tau_x + b' \tau_z\sigma_x + \frac{1}{2}g\mu_B B\sigma_z. 
\label{H_B_line}
\end{equation}
Similar to the point-node case,
the Landau levels can be obtained by replacing $\pi_x \pm i \pi_y$ with $a^\dagger$ and $a$, respectively.
The analytic expression is not available in this case
and here we numerically obtain the Landau levels
by diagonalizing the Hamiltonian matrix with high indexes of $n >100$ truncated.
Figure  \ref{fig_spec_line}(b) presents the obtained Landau levels plotted against $p_z$, 
where we neglect the Zeeman term $g\mu_B B$ for simplicity.
We see that the zero-energy Landau level persists
in the region of $-b' < v p_z < b'$.
It actually corresponds to the region in which
the energy band  Eq.\ (\ref{E_line}) at fixed $p_z$ has point nodes on $p_xp_y$ plane, 
and each of the point nodes accommodates the zero-th Landau level as in graphene.

At $p_z=0$, in particular, the Hamiltonian Eq.\ (\ref{H_B_line}) can be transformed by a certain unitary transformation as
\begin{equation}
H = v \pi_x \sigma_x + (v\pi_y - b' \tau_z) \sigma_y  + \frac 1 2 g\mu_B B\sigma_z,
\end{equation}
which describes a pair of decoupled $2 \times 2$ Dirac cones of $\tau_z=\pm$ sectors.
The Landau levels are obtained as
\begin{align}
\varepsilon_{s n} &= s \sqrt{\Delta_B^2 n+ b^2} \quad (n = 1,2,\cdots),
\nonumber\\
\varepsilon_{0} &= -b,
\label{eq_LL_line_pz0}
\end{align}
where $s=\pm$ and $b=g\mu_B B$. 

\subsection{Lattice model}
\label{sec_lattice}

In addition to the continuum model introduced in the previous section,
we also consider the lattice version of the point-node / line-node semimetals,
to check the validity of  the continuum model in the susceptibility calculation.
For the point-node case,
we introduce a Wilson-Dirac type cubic lattice model,
\begin{align}
H &= u \tau_x (\sigma_x \sin k_x a + \sigma_y \sin k_y a +\sigma_z \sin k_z a )
\nonumber\\
& + [m + r (3-\cos k_x a -\cos k_y a-\cos k_z a)] \tau_z + b\sigma_z,
\label{H_lattice}
\end{align}
where $v p_i (i=x,y,z)$ of Eq.\ (\ref{H}) 
 is simply replaced by $u\sin k_i a$
 with the hopping energy $u$ and the lattice spacing $a$.
The extra term with $r$ is introduced to gap out 
the point nodes other than the origin $(k_x, k_y, k_z)=(0,0,0)$.
In the vicinity of the origin, Eq.\ (\ref{H_lattice}) approximates the continuum version Eq.\ (\ref{H}) 
within the first order of $k_i$.
Figure \ref{fig_spec_lattice}(a) shows the energy band $\varepsilon_{s\mu}(0,p_y,p_z)$ 
of the Dirac semimetal case given by
Eq.\ (\ref{H_lattice}) with $m=b=0$ and $u=r=1$.

In the real space, the Schr\"{o}dinger equation becomes
\begin{align}
E \psi(\Vec{r})
&= [(m+3r)\tau_z + b\sigma_z ]\psi(\Vec{r})
\nonumber\\
&-\frac{r}{2}\tau_z \sum_{i=x,y,z}
[\psi(\Vec{r}+\Vec{a}_i)+ \psi(\Vec{r}-\Vec{a}_i)]
\nonumber\\
&-\frac{i u}{2}\tau_x \sum_{i=x,y,z}
[\psi(\Vec{r}+\Vec{a}_i)- \psi(\Vec{r}-\Vec{a}_i)],
\label{H_lattice_R}
\end{align}
where $\psi(\Vec{r})$ is the four-component wave function
at the lattice point $\Vec{r}$,
and $\Vec{a}_i$ is the unit lattice vector in $i(=x,y,z)$ direction.
The external magnetic field can be included by appending the Peierls phase to the hopping terms,
and also adding the Zeeman energy $g\mu_BB/2$ to $b$.

For the line-node case,
the lattice model corresponding to Eq.\ (\ref{H_line}) becomes
\begin{align}
H &= u \tau_x (\sigma_x \sin k_x a + \sigma_y \sin k_y a +\sigma_z \sin k_z a )
\nonumber\\
& + r (3-\cos k_x a -\cos k_y a-\cos k_z a) \tau_z + b'\tau_z\sigma_x.
\label{H_line_lattice}
\end{align}
Energy spectrum $\varepsilon_{s\mu}(0,p_y,p_z)$ 
with $b'=1.5$ and $u=r=1$ is shown in Fig. \ref{fig_spec_lattice}(b).

  \section{Magnetic Susceptibility}
\label{sec_mag}

The magnetic susceptibility are obtained as in usual manner by differentiating
the thermodynamic potential with the external magnetic field $B$.
Now $B$ appears in the orbital term $\Vec{p}+(e/c)\Vec{A}$ and in the spin Zeeman
term $g\mu_B B$ in the Hamiltonian,
and in the following, we formally treat 
these two magnetic fields for the orbital part and the spin part 
separately as $B_{\rm o}$ and $B_{\rm s}$,
respectively. 

The thermodynamic potential of the whole electronic system is written as
\begin{equation}
\Omega=
-\frac{1}{\beta} \sum_\alpha \varphi(\varepsilon_\alpha)
\label{Omega}
\end{equation}
where
\begin{equation}
\varphi(\varepsilon)=\ln[1+\exp{-\beta(\varepsilon-\zeta)}],
\end{equation}
where $\beta=1/(k_B T)$ with the termperature $T$, 
$\varepsilon_\alpha$ is the energy of the eigenstate $\alpha$,
and $\zeta$ is the chemical potential.
The magnetization is then written as a sum of the spin part and the orbital part,
\begin{align}
M&=
-\frac{\partial\Omega}{\partial B} 
=  -\left(
\frac{\partial\Omega}{\partial B_{\rm s}}+\frac{\partial\Omega}{\partial B_{\rm o}}\right)
\nonumber\\
&\equiv M_{\rm s}+M_{\rm o}.
\end{align}
The magnetic susceptibility is then decomposed as 
\begin{eqnarray}
\chi&=& -\left.\frac{\partial M}{\partial B}\right|_{B=0}
= -\left[
\frac{\partial ^2 \Omega}{\partial B^2_{\rm s}}
+2\frac{\partial ^2 \Omega}{\partial B_{\rm s}\partial B_{\rm o}}
+\frac{\partial ^2\Omega}{\partial B^2_{\rm o}}
\right]_{B_{\rm s}=B_{\rm o}=0}
\nonumber\\
&\equiv&\chi_{\rm s} +\chi_{\rm so}+\chi_{\rm o}.
\label{eq_chi_decomp}
\end{eqnarray}


\begin{figure*}
\begin{center}
 \leavevmode \includegraphics[width=0.9\hsize]{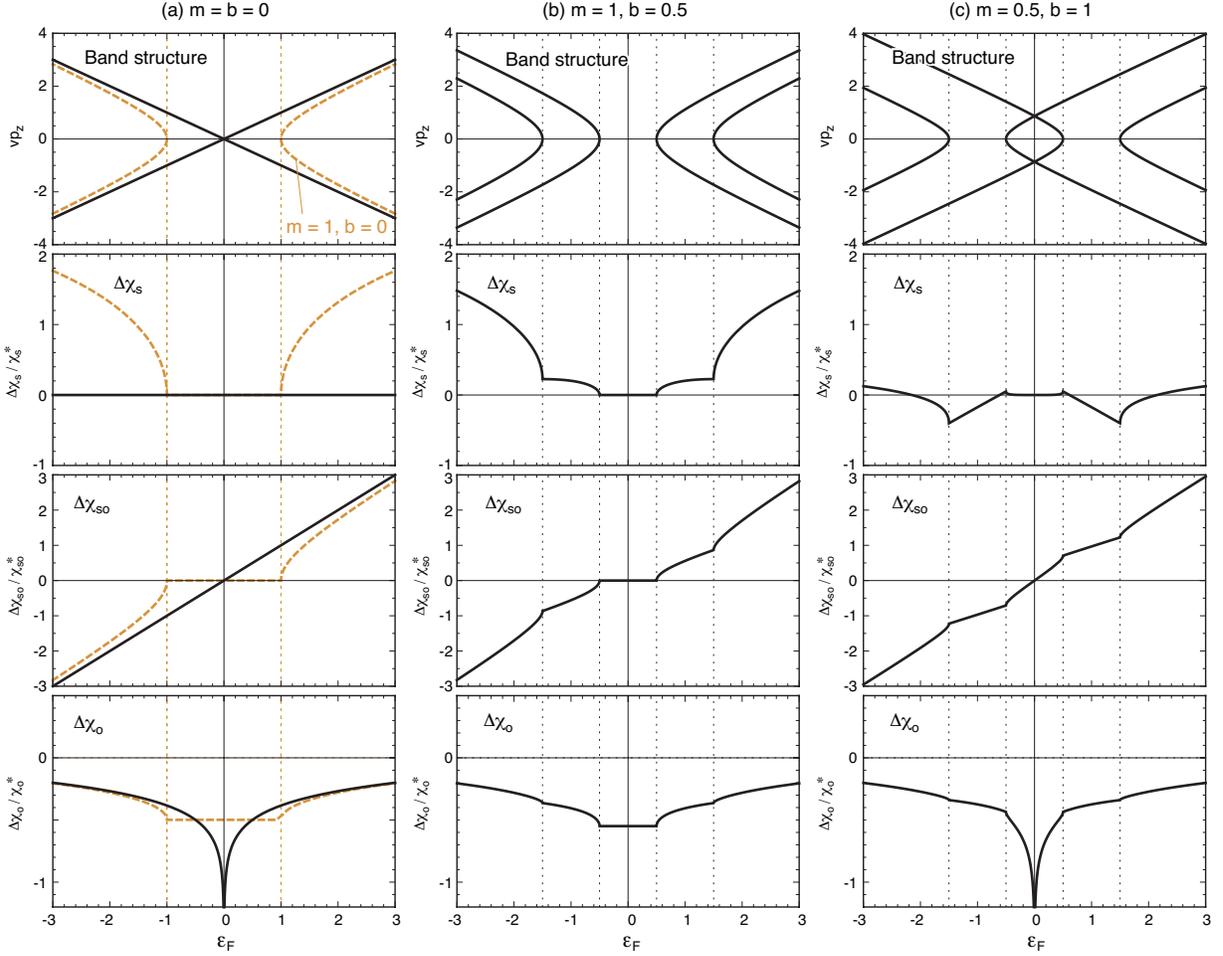}
        \caption{
        Magnetic susceptibility in (a) 
        the Dirac semimetal case ($m= b=0$) and 
        (b) the  semiconducting case ($m=1, b=0.5$)
        and the Weyl semimetal case ($m=0.5, b=1$).
        In each column, the top panel is the 90$^\circ$ rotated band structure $\varepsilon_{s\mu}(0,0,p_z)$
        and the second, third and fourth panels plot $\Delta\chi_{\rm s}$, $\Delta\chi_{\rm so}$ and $\Delta\chi_{\rm o}$
        against the Fermi energy. The dashed curves in (a) are for $m=1, b=0$. 
        }
\label{fig_chi}
\end{center}
\end{figure*}

\begin{figure}
\begin{center}
 \leavevmode \includegraphics[width=0.7\hsize]{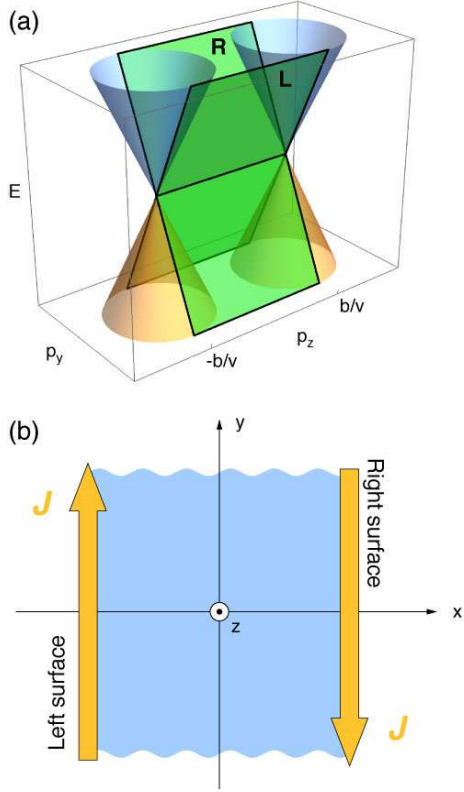}
        \caption{
(a) Surface state bands
in the Weyl semimetal with $m=0$ and a finite $b$. L and R indicate the left and right surface states, respectively.
(b) Corresponding surface current on $xy$ plane.
        }
\label{fig_fermi_arc}
\end{center}
\end{figure}


\subsection{Point-node semimetal}

In the point-node case, the magnetic susceptibility can be derived analytically
using the Landau level spectrum Eq.\ (\ref{eq:landau}). 
The thermodynamic potential per unit volume is explicitly written as
\begin{align}
\Omega =& -\frac{1}{\beta}\frac{1}{2\pi l_B^2} 
\sum_{s\mu p_z}
\sum_{n=n^*}^\infty \varphi[\varepsilon_{s\mu p_z }(x_n)],
\label{thermodynamic}
\end{align}
where  $l_B=\sqrt{c\hbar/(eB_{\rm o})}$ is the magnetic length and we defined
\begin{align}
\sum_{s\mu p_z}=\sum_{s=\pm} \sum_{\mu=\pm} \int_{-\infty}^{\infty} \frac{dp_z}{2\pi \hbar} ,
\end{align}
\begin{align}
\varepsilon_{s\mu p_z}(x_n) &= \varepsilon_{s\mu p_z n} 
\nonumber\\
&\hspace{-15mm} = s \sqrt{m^2+b^2+v^2 p_z^2+x_n+\mu 2b\sqrt{v^2 p_z^2 + m^2}},
\end{align}
\begin{equation}
x_n=n \Delta x, \quad \Delta x = \Delta_B^2 = \frac{2\hbar e v^2}{c}B_{\rm o},
\end{equation}
and 
\begin{eqnarray}
n^*=
\left\{
\begin{array}{ll}
0  & \mbox{for $\tau_{s\mu p_z}=+1$},
\\
1  & \mbox{for $\tau_{s\mu p_z}=-1$}.
\end{array}
\right.
\end{eqnarray}
The $n^*$ specifies the beginning of the Landau level sequence in the sector $(s,\mu,p_z)$,
and it is introduced to remove the non-existing zero-th level argued in Sec.\ \ref{sec_point}.

%

 In Eq.\ (\ref{thermodynamic}), the summation over $n$ can be transformed  
using the Euler-Maclaurin type formula
into a continuous integral over variable $x$ plus a series of the correction terms.
By using a notation $F (x) = \varphi[\varepsilon_{s\mu p_z }(x)]$, we have
\begin{align}
&\sum_{n=n^*}^\infty  F(x_n) 
=
\frac{1}{\Delta x}\Biggl[ \int_0^\infty F(x) dx
\nonumber\\
&\quad
 + \tau_{s\mu p_z}F(0)\frac{\Delta x}{2}-F'(0)\frac{\Delta x^2}{12} + {\cal O}(\Delta x^3)\Biggr],
\end{align}
where the second term in the bracket changes the sign depending on the starting index $n^*$.
%

Noting that $\Delta x \propto B_{\rm o}$, the thermodynamic potential can be expanded 
in powers of $B_{\rm o}$ as
\begin{eqnarray}
\Omega 
&=& \Omega _0 +\lambda _1 B_{\rm o}+\lambda _2 B^2_{\rm o}  + {\cal O}(B^3_{\rm o}),
\end{eqnarray}
where
\begin{eqnarray}
\Omega_0 &=&  -\frac{1}{\beta}\frac{1}{4\pi \hbar^2 v^2}
\sum_{s\mu p_z}
\int_0^\infty \varphi[\varepsilon_{s\mu p_z }(x)] dx,
\label{omega0}
\\
\lambda_1 &=& -\frac{1}{\beta}\frac{1}{4\pi \hbar^2 v^2} 
\frac{\hbar e v^2}{c}
\sum_{s\mu p_z}
\tau_{s\mu p_z}\varphi[\varepsilon_{s\mu p_z }(0)],
\label{lambda1}
\\
\lambda_2 &=& -\frac{1}{\beta}\frac{1}{4\pi \hbar^2 v^2}
\frac{1}{3}
\left(\frac{\hbar e v^2}{c}\right)^2
\sum_{s\mu p_z}
(-1)\frac{\partial}{\partial x}\varphi[\varepsilon_{s\mu p_z }(x)]\Big\rvert_{x=0}.
\nonumber\\
\label{lambda2}
\end{eqnarray}
Finally, the magnetization components are obtained as
\begin{align}
M_{\rm s} 
= - \frac{g\mu_B}{2} \frac{\partial\Omega_0}{\partial b},
\quad M_{\rm o} =  -\lambda_1,
\label{eq_M}
\end{align}
and the susceptibility components as
\begin{align}
&\chi_{\rm s}  = -\left(\frac{g\mu_B}{2}\right)^2  \frac{\partial^2\Omega_0}{\partial b^2},
\quad
\chi_{\rm so}   =  -2 \frac{g\mu_B}{2} \frac{\partial \lambda_1}{\partial b},
\nonumber\\
&\chi_{\rm o}  = -2\lambda_2,
\end{align}
where we used $\partial/\partial B_{\rm s} =(g\mu_B/2) \partial/\partial b$.


By using the explicit form of $\varepsilon_{s\mu p_z}(x)$,
we end up with the formulas,
\begin{align}
&\chi_{\rm s}
=\frac{1}{4\pi\hbar ^2v^2}\left(\frac{g\mu_B}{2}\right)^2
\sum_{s\mu p_z} 
\int_0^\infty dx 
\nonumber\\
&\hspace{15mm}
\Biggl[
-f'(\varepsilon)
 \left(\frac{\partial \varepsilon}{\partial b}\right)^2+
 f(\varepsilon)
 \left( - \frac{\partial ^2 \varepsilon}{\partial b^2}\right)
 \Biggr],
 \label{chi_s}
 \\
 &\chi_{\rm so}
=\frac{1}{4\pi\hbar ^2v^2} \frac{\hbar e v^2}{c} g\mu_B 
\sum_{s\mu p_z} f[\varepsilon_{s\mu p_z }(0)],
 \label{chi_so}
 \\
& \chi_{\rm o}
= -\frac{1}{4\pi\hbar ^2v^2}
\frac{1}{3}\left(\frac{\hbar e v^2}{c}\right)^2
\sum_{s\mu p_z}
\frac{f[\varepsilon_{s\mu p_z }(0)]}{\varepsilon_{s\mu p_z }(0)},
\label{chi_o}
\end{align}
where $\varepsilon=\varepsilon_{s\mu p_z }(x)$, $f(\varepsilon)=\left[1+e^{\beta(\varepsilon-\zeta)}\right]^{-1}$
is the Fermi distribution function,
and we used the relation $\varphi'(\varepsilon)=-\beta f(\varepsilon)$
and Eq.\ (\ref{eq_zeroLL}).
The variable $\zeta$ in $f(\varepsilon)$ is the chemical potential, and 
it will be denoted by $\varepsilon_F$ in the limit of $T=0$.

In the spin component $\chi_{\rm s}$, the first term in the bracket 
is contributed from the states at the Fermi surface,
and it corresponds to the conventional Pauli paramagnetism.
The second term depends on all the occupied states below the Fermi energy,
and it describes the Van-Vleck paramagnetism.
The spin-orbit component $\chi_{\rm so}$, Eq.\ (\ref{chi_so}), 
is formally proportional to the number of states 
below the Fermi energy  in the zero-th level $\varepsilon_{s\mu p_z }(0)$,
including both existing $(\tau_{s\mu p_z}=+1)$ or non-existing $(-1)$ branches.
 %
For the orbital part $\chi_{\rm o}$, Eq.\ (\ref{chi_o}),  
the summation in the index $s$ gives in the limit of $T\to 0$,
\begin{align}
\sum_{s=\pm}
\frac{f[\varepsilon_{s\mu p_z }(0)]}{\varepsilon_{s\mu p_z }(0)}
=
\frac{\theta(|\varepsilon_F| - |\varepsilon_{\mu p_z 0}|)}{|\varepsilon_{\mu p_z 0}|},
\end{align}
where $\theta(x)$ is the step function returning 1(0) for $x>0(<0)$,
and we used the relation $\varepsilon_{s\mu p_z }(0) = s |\varepsilon_{\mu p_z 0}|$.
Considering the minus sign in front of Eq.\ (\ref{chi_o}),
each sector labeled by $(\mu, p_z)$ contributes to the diamagnetic susceptibility 
in the energy range $-|\varepsilon_{\mu p_z 0}| < \varepsilon_F < |\varepsilon_{\mu p_z 0}|$,
and its amplitude is inversely proportional to the width of the energy window. 
The delta function diamagnetism in graphene is obtained in the limit of $\varepsilon_{\mu p_z 0}\to 0$. 
\cite{koshino2010anomalous}

Now let us apply the above susceptibility formula to several different situations with specific $(m,b)$'s. 
When calculating $\chi$, we introduce a momentum cut-off in the integral as $|p_z| < \varepsilon_c/v$
and $x < \varepsilon_c^2$ with a sufficiently large energy $\varepsilon_c$,
to avoid the divergence of the integral and also
to simulate the finite-sized Brillouin zone 
in real materials.  As a result, the calculated susceptibility is shown to 
include a constant term (i.e., independent of the Fermi energy) 
which explicitly depends on the cut-off $\varepsilon_c$. 
In the real material, however, such an offset term 
should be determined by the whole band structure beyond the description
of the present linear Hamiltonian, \cite{ominato2013orbital}
and it is not determinate in the continuum scheme.
In the following, therefore, we argue about only the relative susceptibility as a function of the Fermi energy, 
neglecting the offset term.
The problem of the offset term will be addressed later in this section using the lattice model.

We first consider the Dirac semimetal case $(m=b=0)$.
The susceptibility at $T=0$ is explicitly calculated as functions of the Fermi energy $\varepsilon_F$ as
\begin{eqnarray}
&& \Delta\chi_{\rm s} =  0,
\nonumber \\
 && \Delta\chi_{\rm so}  = \frac{e^2v}{\pi^2 \hbar c^2} 
  \frac{g\mu_B c}{2\hbar e v^2} 
  \varepsilon_F,
 \label{eq_chi_b0m0}
\\
 && \Delta\chi_{\rm o} = -\frac{e^2v}{12\pi^2 \hbar c^2} 
\ln \frac{\varepsilon_c}{|\varepsilon_F|},
 \nonumber
\end{eqnarray}
where $\Delta \chi_{\rm s}$ etc. represents the relative susceptibility with 
the constant term appropriately chosen.
The three susceptibility components are plotted in separate panels in Fig.\ \ref{fig_chi}(a),
together with the 90$^\circ$-rotated band structure in the top panel.
The orbital diamagnetism $\Delta\chi_{\rm o}$ logarithmically diverges in the diamagnetic direction
at the band touching point. \cite{koshino2010anomalous}
The spin term $\Delta\chi_{\rm s}$ becomes completely Fermi-energy independent,
where the energy dependences of Pauli term and Van-Vleck term completely cancel out.
The spin-orbital term $\Delta\chi_{\rm so}$ is a simple linear function
monotonically rising in increasing $\varepsilon_F$.

The vanishing $\Delta \chi_{\rm s}$ can be clearly understood
by considering the process to increase the spin Zeeman term $b$.
The Hamiltonian  Eq.\ (\ref{H}) with $m=0$ is
decoupled to $\tau_x=\pm$ sectors as
\begin{equation}
H_{\pm} = \pm v [\sigma_x p_x + \sigma_y p_y + \sigma_z (p_z \mp b/v)],
\end{equation}
which describes the Weyl semimetal composed of 
independent Weyl cones centered at $\Vec{p}=(0,0,\pm b/v)$. 
The spin field $b$ just horizontally separates the 
Dirac cones in the wave-space, so that the spin density never changes in this process.

On the other-hand, the finite $\Delta \chi_{\rm so}$ indicates 
that the finite orbital magnetization $M_{\rm o}$ is induced by increasing $b$,
and this is explicitly calculated as
\begin{equation}
M_{\rm o} = \frac{1}{2} \Delta\chi_{\rm so}\, \frac{b}{g\mu_B/2} 
=\frac{e}{2\pi^2\hbar^2v c} \varepsilon_F b.
\label{eq_M_o}
\end{equation}
The physical origin of the induced $M_{\rm o}$
becomes clear by introducing the surface to the system.
When we consider the slab geometry which is finite in $x$-direction
while infinite in $y$ and $z$,  we have a pair of surface localized bands
which connect the two Weyl cones as shown in Fig.\ \ref{fig_fermi_arc}(a). \cite{wan2011topological,burkov2011weyl}
The surface states then give the counter flows parallel to $y$ directions on the opposite surfaces
as illustrated in Fig.\ \ref{fig_fermi_arc}(b),
and they contribute to a magnetization in $z$ direction.
The velocity of the surface state along $y$ is equal to $v$.
The area density of the surface electrons existing between the zero energy to the Fermi energy
 is $n_{\rm surf}=(\varepsilon_F /v) (2b/v)/(2\pi\hbar)^2$ by considering the corresponding momentum space.
The total surface current density then becomes $j = e v n_{\rm surf} = e\varepsilon_F b/(2\pi^2\hbar^2 v)$. 
The magnetization induced by the surface current, $j/c$, 
exactly coincide with Eq.\ (\ref{eq_M_o}).



The susceptibility calculation can be extended to the general cases with a finite mass $m$.
At $b=0$, in paticular, we have
\begin{eqnarray}
&& \Delta\chi_{\rm s} 
= \frac{e^2v}{\pi^2 \hbar c^2} 
 \left(\frac{g\mu_B c}{2\hbar e v^2}\right)^2 
 \nonumber\\
&&\hspace{20mm} \times
m^2 \theta(|\varepsilon_F|-m) \ln \frac{\sqrt{\varepsilon_F^2-m^2}+|\varepsilon_F|}{m},
\\
 && \Delta\chi_{\rm so}
 = \frac{e^2v}{\pi^2 \hbar c^2}  
 \frac{g\mu_B c}{2\hbar e v^2} 
\nonumber\\
&&\hspace{20mm} \times
{\rm sgn}(\varepsilon_F)  \theta(|\varepsilon_F|-m) 
\sqrt{\varepsilon_F^2-m^2},
 \label{eq_chi_b0}
\\
 && \Delta\chi_{\rm o}
  = -\frac{e^2v}{12\pi^2 \hbar c^2} 
 \times \left\{
\begin{array}{ll}
\displaystyle \ln \frac{2 \varepsilon_c}{m} & |\varepsilon_F|< m,
\\ 
\displaystyle \ln \frac{2 \varepsilon_c}{\sqrt{\varepsilon_F^2-m^2}+|\varepsilon_F|} & |\varepsilon_F|> m, 
\end{array}
 \right.
 \nonumber\\
\end{eqnarray}
which are plotted as dashed curves in Fig.\ \ref{fig_chi}(a) ($m=1, b=0$).
In $\Delta\chi_{\rm s}$, the cancelation of Pauli term and Van-Vleck term 
becomes incomplete, leaving a positive component inside the band.
We also see the log peak of $\Delta\chi_{\rm o}$ is truncated,
and the increase of $\Delta\chi_{\rm so}$ is interrupted by the energy gap.
Now the system is in the trivial phase without the surface states 
so $\Delta\chi_{\rm so}$ is contributed by the bulk states.


The susceptibility formula for the general $(m,b)$ is complicated 
and the full expression is presented in Appendix \ref{sec_app}. 
Here we plot some representative cases in Fig.\ \ref{fig_chi}(b) and (c).
When we introduce $b$ smaller than $m$ [Fig.\ \ref{fig_chi}(b) ($m=1, b=0.5$)], 
a band spit gives minor changes to the susceptibility curve of $m=1, b=0$ in Fig.\ \ref{fig_chi}(a).
When $b$ is larger than $m$,
the system enters the Weyl semimetal phase  [Fig.\ \ref{fig_chi}(c) ($m=0.5, b=1$)], 
and we observe a similar behavior to the Dirac semimetal case
in the low-energy region reflecting the emergence of the point nodes;
i.e., the vanishing  $\Delta \chi_{\rm s}$, the linear increase of $\Delta \chi_{\rm so}$
and log divergence of $\Delta \chi_{\rm o}$.
Outside the linear band region,
$\Delta \chi_{\rm s}$ becomes negative (i.e., relatively diamagnetic compared to the zero energy) 
near the second band edges $\varepsilon_F=\pm(m+b)$,
and it increases again outside.
We notice that $\Delta\chi_{\rm so}$ is always an odd function of $\varepsilon_F$
while other two components, $\chi_{\rm s}$ and $\chi_{\rm o}$,
are even functions. This is closely related to the fact that
the electron-hole symmetry of the energy spectrum 
is broken solely by the zero-th Landau level 
when $B_{\rm o}$ and $B_{\rm s}$ coexist.


In Fig.\ \ref{fig_chi}, the three susceptibility components are plotted
in different characteristic scales,
\begin{eqnarray}
&& \chi^*_{\rm s} 
= \frac{e^2v}{\pi^2 \hbar c^2}\,  \eta^2,
\quad
 \chi^*_{\rm so}
 = \frac{e^2v}{\pi^2 \hbar c^2}  \, \eta,
\quad
\chi^*_{\rm o}
  = \frac{e^2v}{\pi^2 \hbar c^2}, 
\label{chi_units}
\end{eqnarray} 
where 
\begin{equation}
\eta =  \frac{g\mu_B c}{2\hbar e v^2} \varepsilon_0,
\end{equation}
and $\varepsilon_0$ is the energy unit to measure $m$, $b$ and $\varepsilon_F$.
The relative magnitudes of three components are determined by the dimensionless factor $\eta$,
which is the ratio of the spin magnetic moment $g\mu_B/2$
to the characteristic orbital magnetic moment  $\mu_0 = \hbar e v^2/(\varepsilon_0 c)$
associated with the energy scale $\varepsilon_0$.
In the typical velocity operator $v$ of the order of  $10^5$ m/s, for instance,
$\eta$ becomes the order of $\varepsilon_0/(100 {\rm meV})$. 
When the typical energy scales for the band structure
such as $\varepsilon_F$ and $m$ are much smaller than 100 meV, 
$\chi_{\rm o}$ is dominant, 
$\chi_{\rm so}$ is in the middle range,
and $\chi_{\rm s}$ is the smallest.
In this case, the system should be diamagnetic in total.


Finally, we calculate the susceptibility of the corresponding lattice model Eq.\ (\ref{H_lattice})
to check the validity of the continuum approximation.
Here the computations are done numerically;
we first obtain the eigenenergies of the Hamiltonian at the several discrete points
of $(B_{\rm orb}, B_{\rm spin})$ which are small enough, and compute the thermodynamic potential $\Omega$
at each point.
Then we take the derivative of $\Omega$ using the differential approximation, and obtain the susceptibility
using Eq.\ (\ref{eq_chi_decomp}).
In the calculation, we assume a finite temperature ($k_BT=0.1$)
to smear out the discrete level structure at finite $B_{\rm orb}$.

\begin{figure}
\begin{center}
 \leavevmode \includegraphics[width=0.8\hsize]{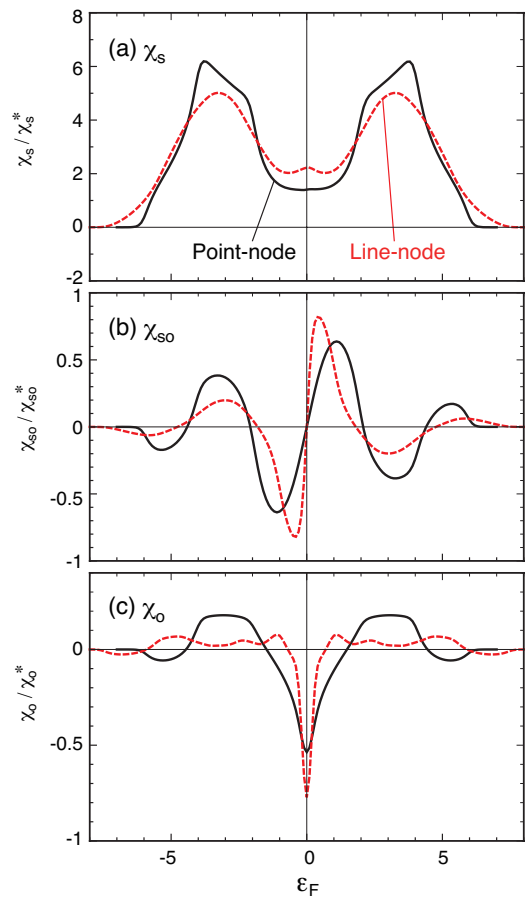}
        \caption{
       Susceptibility (a)  $\chi_{\rm s}$, (b) $\chi_{\rm so}$ and (c) $\chi_{\rm o}$
         calculated for (a) the Dirac semimetal [Eq.\ (\ref{H_lattice}) with $m=b=0$ and $u=r=1$] 
         and (b) the line-node semimetal [Eq.\ (\ref{H_line_lattice}) with  $b'=1.5$ and $u=r=1$] .
        }
\label{fig_chi_lattice}
\end{center}
\end{figure}

The results for the Dirac semimetal  [$m=b=0$ and $u=r=1$]
are shown as solid curves in Fig.\ \ref{fig_chi_lattice}, where (a), (b) and (c) 
plot $\chi_{\rm s}$, $\chi_{\rm so}$ and $\chi_{\rm o}$, respectively.
They are the absolute $\chi$'s and not the relative $\Delta \chi$'s.
The vertical axis is scaled by Eq.\ (\ref{chi_units}) with $v$ replaced with the corresponding quantity, $ua/\hbar$.
In the low-energy region $|\varepsilon_F|<1$,
we actually see that the energy dependence of each $\chi$ well resembles the continuum counterpart
in Fig.\ \ref{fig_chi}(a), including the flat dependence of $\chi_{\rm s}$,
the linear behavior $\chi_{\rm so}$ and the negative log peak of  $\chi_{\rm o}$.
We see that $\chi_{\rm s}$ and $\chi_{\rm o}$ are even function,
and both have the paramagnetic offsets compared to the continuum model.
In contrast, $\chi_{\rm so}$ is an odd function and it vanishes at $\varepsilon_F=0$.
The results clearly show that the continuum calculation 
correctly describes the Fermi-energy dependence of the susceptibility in the low-energy region.

\begin{figure}
\begin{center}
 \leavevmode \includegraphics[width=0.7\hsize]{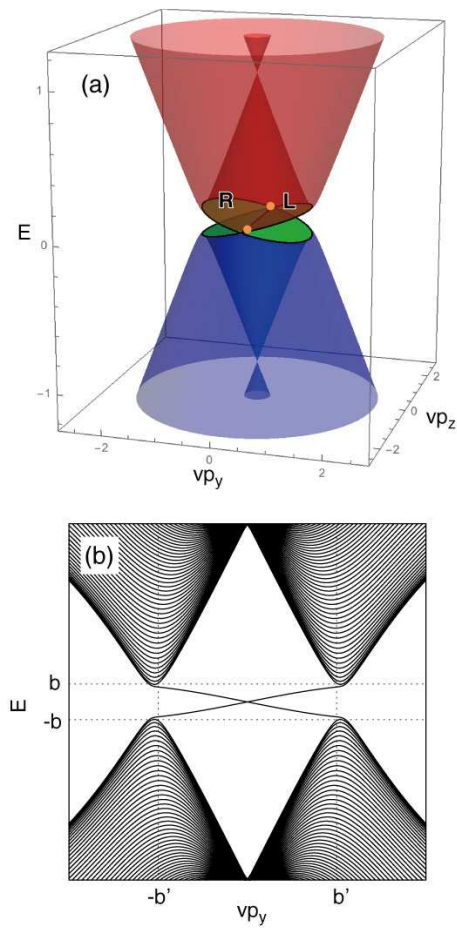}
        \caption{
        (a) Energy spectrum of the line node Hamiltonian Eq.\ (\ref{H_line}) with $b\sigma_z$ added
        ($b'=1, b=0.1$). The green parts represent the surface state bands on the left (L) and right (R) surfaces parallel to
        $yz$ plane.
        (b)  Band structure of the corresponding lattice model [Eq.\ (\ref{H_line_lattice}) with  $b'=1$, $b=0.1$ and $u=r=1$]
        with the slab geometry of the width 100 sites in $x$ direction.
        The energy bands are plotted against $p_y$, where $p_z$ is fixed to 0.
        }
\label{fig_line_ribbon}
\end{center}
\end{figure}


\subsection{Line-node semimetal}

For the line node semimetal Eq.\ (\ref{H_line}),
the susceptibility can be estimated in an approximate manner as follows.
As argued in Sec. \ref{sec_lattice}, 
the energy band are degenerate along the circle node on $p_yp_z$-plane.
If we regard $p_z$ as a parameter and view the Hamiltonian
as a two-dimensional system on $xy$-plane,
the energy spectrum on the $p_xp_y$-space contains a pair
of Weyl nodes when $-b'/v < p_z < b'/v$.
At $p_z=0$, in particular, the Hamiltonian can be separated into independent $2\times 2$ Weyl Hamiltonians, where
we have the analytic expression of the Landau levels, Eq.\ (\ref{eq_LL_line_pz0}).
Since the Landau level spectrum near the zero energy does not rapidly change when $p_z$ moves away from 0
as shown in Fig.\ \ref{fig_spec_line}(b),
we can roughly approximate
the whole system by a simplified model
in which the energy spectrum at $p_z=0$ just persists throughout the region of $-b'/v < p_z < b'/v$.

Then the magnetic susceptibility is immediately calculated by integrating
the contribution of $p_z=0$ through $-b'/v < p_z < b'/v$.
Specifically,  
we substitute Eqs.\ (\ref{chi_s}), (\ref{chi_so}) and (\ref{chi_o}) with
\begin{align}
&\varepsilon_{s\mu p_z}(x_n) = s \sqrt{x_n+ b^2},
\nonumber\\
&
\tau_{s\mu p_z}= -s,
\label{eq_simple_line_1}
\end{align}
and 
\begin{align}
\sum_{s \mu p_z}= \sum_{s=\pm} \sum_{\mu=\pm}
 \int_{b'/v}^{-b'/v} 
 \frac{dp_z}{2\pi \hbar},
 \label{eq_simple_line_2}
\end{align}
where $\mu$ labels the two identical Weyl nodes at fixed $p_z$.
As a result, we obtain
\begin{eqnarray}
&& \Delta\chi_{\rm s} 
= \frac{e^2v}{\pi^2 \hbar c^2} 
 \left(\frac{g\mu_B c}{2\hbar e v^2}\right)^2 
 2b' (-|\varepsilon_F|)
\nonumber\\
 && \Delta\chi_{\rm so}
 = \frac{e^2v}{\pi^2 \hbar c^2} 
  \frac{g\mu_B c}{2\hbar e v^2} 
\, 2b' \theta(\varepsilon_F)
\nonumber\\
 && \Delta\chi_{\rm o}
  = -\frac{e^2v}{3\pi^2 \hbar c^2} \, b' \delta(\varepsilon_F).
\label{chi_line_approx}
\end{eqnarray}
Now the orbital term $\Delta\chi_{\rm o}$ has a negative delta-function as in graphene,
and this is natural because the system is a sum of 2D gapless systems over a finite range of $p_z$.
The singularity of the diamagnetism is therefore much stronger than in the log peak in the point-node case.
$\Delta\chi_{\rm so}$ becomes a step function which has a discontinuity at zero energy.
 As argued in the previous section, $\Delta\chi_{\rm so}$ is formally given by
the number of states in the zero-th level below the Fermi energy,
and now it jumps at $\varepsilon_F=0$ because all the zero energy level lies there.
The spin magnetism $\Delta\chi_{\rm s}$ linearly decreases as the Fermi energy goes away from zero.

The line-node is generally not robust unlike the point-node in the Weyl semimetal,
and it can be gapped out by a perturbation. \cite{burkov2011topological}
But still, the delta-function diamagnetic peak never just disappears
but widens to the gap region keeping the total area [i.e. integral of $\chi(\varepsilon_F)]$,
as shown for the gapped 2D case. \cite{koshino2010anomalous}


We also calculate the susceptibility for
the corresponding lattice model, Eq.\ (\ref{H_line_lattice}).
The results are shown as red dashed curves in Fig.\ \ref{fig_spec_line},
where we take $u=r=1$ and $b'=3$.
As seen from the spectrum in Fig.\ \ref{fig_spec_line},
the continuum model is valid only in a small energy range $|\varepsilon_F|<0.5$,
and the susceptibility inside this region
qualitatively agrees with the approximate expression Eq.\ (\ref{chi_line_approx}).
Specifically, we see a sharp diamagnetic peak in $\chi_{\rm o}$
as well as the abrupt rise in $\chi_{\rm so}$,
while the features are smeared by the finite temperature $k_BT = 0.1$.
We also see a small cusp in $\chi_{\rm s}$, which corresponds to the
linear decrease in Eq.\ (\ref{chi_line_approx}).


The step-like increase of $\chi_{\rm so}$  at zero Fermi energy 
can be again understood in terms of the topological surface state. 
Similarly to the point-node case,
the finite $\Delta \chi_{\rm so}$ indicates 
that the orbital magnetization $M_{\rm o} = \Delta \chi_{\rm so} b/(g\mu_B)$
is induced by the spin Zeeman field $b$. 
The step-like jump in $\Delta \chi_{\rm so}$ in Eq.\ (\ref{chi_line_approx}) then
corresponds to magnetization accumulated near the zero energy, of which amount is
\begin{equation}
M_{\rm o} = \frac{eb'b}{\pi^2\hbar^2v c}.
\label{eq_M_o_line}
\end{equation}

If we add the spin field $b\sigma_z$ to the Hamiltonian Eq.\ (\ref{H_line}),
the line node is gapped leaving two point nodes at $(p_y,p_z)=(0,\pm b'/v)$
as illustrated in Fig.\ \ref{fig_line_ribbon}(a).
In a system bound by the surfaces perpendicular to 
 $x$-direction, we have the left and right surface bands to connect the point nodes,
which are indicated by two green ovals in Fig.\ \ref{fig_line_ribbon}(a).
The actual band structure at $p_z=0$ in the corresponding
lattice model with the slab geometry 
is shown Fig.\ \ref{fig_line_ribbon}(b),
where we see that a pair of surface localized bands.
The surface bands disperse in $p_y$ direction and the velocity is given by $v_{\rm surf} = (b/b')v$.
Each surface state contributes to the electric current 
$\pm e v_{\rm surf}$ counter-flowing the opposite surfaces.
If we adopt the simplification used above
replacing the spectrum of $-b'/v < p_z < b'/v$ with $p_z=0$,
the area density of the surface electrons  is $n_{\rm surf}=(2b'/v)^2/(2\pi\hbar)^2$
by considering the momentum region  $-b'/v < p_y, p_z < b'/v$.
Finally, the total surface current density is 
$j = e v_{\rm surf} n_{\rm surf} = ebb'/(\pi^2\hbar^2v)$,
and the total magnetization given by the surface current
then becomes $M_{\rm 0} = j/c = eb'b / (\pi^2\hbar^2v c)$,
which exactly coincides with Eq.\ (\ref{eq_M_o_line}).
A similar orbital magnetization induced by the spin Zeeman field
was also argued in the surface Dirac cone in the topological insulator
gapped by the magnetic perturbation. \cite{nomura2010electric, tserkovnyak2015spin}

\section{Conclusion}
\label{sec_concl}

We studied the response to the external magnetic field in various three-dimensional nodal systems.
We showed that the magnetic susceptibility consists of the orbital term, the spin term and the spin-orbit cross term,
and they have different magnitudes and different  characteristic properties.
The orbital term is diamagnetic near the band touching point,
and it exhibits a log divergence at the point node while a delta-function divergence at the line node.
The spin-orbit cross term is caused by the spin-orbit interaction,
and in the nodal semimetals it is closely related to the chiral surface current carried by
the topological surface states. In the point-node case, it is a linear function in the Fermi energy
while in line node case, it discontinuously jumps right at the line-node. 
The spin susceptibility includes the Pauli and Van-Vleck paramagnetism,
and in the point node case it is found to be completely Fermi-energy independent 
in the linear band regime near the degenerate points.
For the calculation,  we consider both continuum linear model and the corresponding lattice model,
to show that the continuum model correctly describes the Fermi-energy dependence of 
the low-energy susceptibility except for the constant offset term.

The authors thank Kentaro Nomura for helpful discussions.
This work was supported by Grants-in- Aid for Scientific research (Grants No. 25107005). 

\appendix

\section{Susceptibility in general point-node semimetals}
\label{sec_app}

For the Hamiltonian Eq.\ (\ref{H}) with general $m$ and $b$,
the susceptibility is explicitly calculated using 
Eqs.\ (\ref{chi_s}), (\ref{chi_so}) and (\ref{chi_o}) as,
\begin{align}
\Delta\chi_{\rm s} 
=& \frac{e^2v}{\pi^2 \hbar c^2} 
 \left(\frac{g\mu_B c}{2\hbar e v^2}\right)^2 
\sum_{s,\mu} \theta[s(\varepsilon_F-\mu b) -|m|]
 \nonumber\\
&\hspace{0mm} \times
\left[
P(t_\mu) +
s\left(-\frac{1}{2}\varepsilon_F+\mu b\right)t_\mu
\right],
\\
\Delta\chi_{\rm so}
 =& \frac{e^2v}{\pi^2 \hbar c^2}  \frac{g\mu_B c}{2\hbar e v^2}
 \sum_{s,\mu} 
 \frac{s}{2}\theta[s(\varepsilon_F-\mu b) -|m|] t_\mu,
 \\
  \Delta\chi_{\rm o} 
  =& \frac{e^2v}{12\pi^2 \hbar c^2} 
  \Bigl[
  Q_-(|b|-|\varepsilon_F|)\theta(|b|-|m|-|\varepsilon_F|)
  \nonumber\\
&+ \sum_{\mu} Q_\mu (|\varepsilon_F|-\mu|b|)\theta(|\varepsilon_F|-|m|-\mu|b|)
\Bigr],
 \end{align}
 where $s=\pm$, $\mu = \pm$, and
\begin{align}
&t_\mu = \sqrt{(\varepsilon_F-\mu b)^2-m^2},
\nonumber\\
&P(t) 
=\frac{1}{2}\left(
t\sqrt{t^2+m^2} + m^2 \log \frac{t+\sqrt{t^2+m^2}}{|m|}
\right),
\nonumber\\
&Q_\mu (t) 
=
\left\{
\renewcommand{\arraystretch}{1.7} 
\begin{array}{l}
\displaystyle
{\rm arccosh}\frac{t}{m}-\frac{\mu |b|}{\sqrt{m^2-b^2}}
{\rm arccos}\frac{m^2+\mu|b| t}{m(t+\mu|b|)}
\\
\hspace{50mm} (|m|>|b|),
\\
\displaystyle
{\rm arccosh}\frac{t}{m}-\frac{|b|}{\sqrt{b^2-m^2}}
{\rm arccosh}
\left|
\frac{m^2+\mu|b| t}{m(t+\mu|b|)}
\right|
\\
\hspace{50mm} (|m|<|b|).
\end{array}
\right.
\renewcommand{\arraystretch}{1.} 
\end{align}
The magnetization Eq.\ (\ref{eq_M}) can also be calculated as
\begin{align}
M_{\rm s} 
=& \frac{e}{2\pi^2 \hbar^2 v c} 
\frac{g\mu_B c}{2\hbar e v^2} 
\sum_{s,\mu} \theta[s(\varepsilon_F-\mu b) -|m|]
 \nonumber\\
& \times
\Bigl\{
(-\mu \varepsilon_F + 2b)P(t_\mu)
+s[ -\varepsilon_F b +\mu (b^2+m^2) ] t_\mu
 \nonumber\\
& \qquad
+ \frac{1}{3}s\mu t_\mu^3
\Bigr\},
\\
M_{\rm o}
 =& 
 \frac{e}{4\pi^2 \hbar v c} 
\sum_{s,\mu} 
\theta[s(\varepsilon_F-\mu b) -|m|]
 \nonumber\\
&\hspace{0mm} \times
\mu \left[
P(t_\mu)- s (\varepsilon_F-\mu b) t_\mu
\right].
 \end{align}

\bibliography{weyl_mag}

\begin{thebibliography}{38}%
\makeatletter
\providecommand \@ifxundefined [1]{%
 \@ifx{#1\undefined}
}%
\providecommand \@ifnum [1]{%
 \ifnum #1\expandafter \@firstoftwo
 \else \expandafter \@secondoftwo
 \fi
}%
\providecommand \@ifx [1]{%
 \ifx #1\expandafter \@firstoftwo
 \else \expandafter \@secondoftwo
 \fi
}%
\providecommand \natexlab [1]{#1}%
\providecommand \enquote  [1]{``#1''}%
\providecommand \bibnamefont  [1]{#1}%
\providecommand \bibfnamefont [1]{#1}%
\providecommand \citenamefont [1]{#1}%
\providecommand \href@noop [0]{\@secondoftwo}%
\providecommand \href [0]{\begingroup \@sanitize@url \@href}%
\providecommand \@href[1]{\@@startlink{#1}\@@href}%
\providecommand \@@href[1]{\endgroup#1\@@endlink}%
\providecommand \@sanitize@url [0]{\catcode `\\12\catcode `\$12\catcode
  `\&12\catcode `\#12\catcode `\^12\catcode `\_12\catcode `\%12\relax}%
\providecommand \@@startlink[1]{}%
\providecommand \@@endlink[0]{}%
\providecommand \url  [0]{\begingroup\@sanitize@url \@url }%
\providecommand \@url [1]{\endgroup\@href {#1}{\urlprefix }}%
\providecommand \urlprefix  [0]{URL }%
\providecommand \Eprint [0]{\href }%
\providecommand \doibase [0]{http://dx.doi.org/}%
\providecommand \selectlanguage [0]{\@gobble}%
\providecommand \bibinfo  [0]{\@secondoftwo}%
\providecommand \bibfield  [0]{\@secondoftwo}%
\providecommand \translation [1]{[#1]}%
\providecommand \BibitemOpen [0]{}%
\providecommand \bibitemStop [0]{}%
\providecommand \bibitemNoStop [0]{.\EOS\space}%
\providecommand \EOS [0]{\spacefactor3000\relax}%
\providecommand \BibitemShut  [1]{\csname bibitem#1\endcsname}%
\let\auto@bib@innerbib\@empty
\bibitem [{\citenamefont {Murakami}(2007)}]{murakami2007phase}%
  \BibitemOpen
  \bibfield  {author} {\bibinfo {author} {\bibfnamefont {S.}~\bibnamefont
  {Murakami}},\ }\href@noop {} {\bibfield  {journal} {\bibinfo  {journal} {New
  J. Phys.}\ }\textbf {\bibinfo {volume} {9}},\ \bibinfo {pages} {356}
  (\bibinfo {year} {2007})}\BibitemShut {NoStop}%
\bibitem [{\citenamefont {Young}\ \emph {et~al.}(2012)\citenamefont {Young},
  \citenamefont {Zaheer}, \citenamefont {Teo}, \citenamefont {Kane},
  \citenamefont {Mele},\ and\ \citenamefont {Rappe}}]{young2012dirac}%
  \BibitemOpen
  \bibfield  {author} {\bibinfo {author} {\bibfnamefont {S.~M.}\ \bibnamefont
  {Young}}, \bibinfo {author} {\bibfnamefont {S.}~\bibnamefont {Zaheer}},
  \bibinfo {author} {\bibfnamefont {J.~C.}\ \bibnamefont {Teo}}, \bibinfo
  {author} {\bibfnamefont {C.~L.}\ \bibnamefont {Kane}}, \bibinfo {author}
  {\bibfnamefont {E.~J.}\ \bibnamefont {Mele}}, \ and\ \bibinfo {author}
  {\bibfnamefont {A.~M.}\ \bibnamefont {Rappe}},\ }\href@noop {} {\bibfield
  {journal} {\bibinfo  {journal} {Phys. Rev. Lett.}\ }\textbf {\bibinfo
  {volume} {108}},\ \bibinfo {pages} {140405} (\bibinfo {year}
  {2012})}\BibitemShut {NoStop}%
\bibitem [{\citenamefont {Wang}\ \emph {et~al.}(2012)\citenamefont {Wang},
  \citenamefont {Sun}, \citenamefont {Chen}, \citenamefont {Franchini},
  \citenamefont {Xu}, \citenamefont {Weng}, \citenamefont {Dai},\ and\
  \citenamefont {Fang}}]{wang2012dirac}%
  \BibitemOpen
  \bibfield  {author} {\bibinfo {author} {\bibfnamefont {Z.}~\bibnamefont
  {Wang}}, \bibinfo {author} {\bibfnamefont {Y.}~\bibnamefont {Sun}}, \bibinfo
  {author} {\bibfnamefont {X.-Q.}\ \bibnamefont {Chen}}, \bibinfo {author}
  {\bibfnamefont {C.}~\bibnamefont {Franchini}}, \bibinfo {author}
  {\bibfnamefont {G.}~\bibnamefont {Xu}}, \bibinfo {author} {\bibfnamefont
  {H.}~\bibnamefont {Weng}}, \bibinfo {author} {\bibfnamefont {X.}~\bibnamefont
  {Dai}}, \ and\ \bibinfo {author} {\bibfnamefont {Z.}~\bibnamefont {Fang}},\
  }\href@noop {} {\bibfield  {journal} {\bibinfo  {journal} {Phys. Rev. B}\
  }\textbf {\bibinfo {volume} {85}},\ \bibinfo {pages} {195320} (\bibinfo
  {year} {2012})}\BibitemShut {NoStop}%
\bibitem [{\citenamefont {Burkov}\ and\ \citenamefont
  {Balents}(2011)}]{burkov2011weyl}%
  \BibitemOpen
  \bibfield  {author} {\bibinfo {author} {\bibfnamefont {A.}~\bibnamefont
  {Burkov}}\ and\ \bibinfo {author} {\bibfnamefont {L.}~\bibnamefont
  {Balents}},\ }\href@noop {} {\bibfield  {journal} {\bibinfo  {journal} {Phys.
  Rev. Lett.}\ }\textbf {\bibinfo {volume} {107}},\ \bibinfo {pages} {127205}
  (\bibinfo {year} {2011})}\BibitemShut {NoStop}%
\bibitem [{\citenamefont {Burkov}\ \emph {et~al.}(2011)\citenamefont {Burkov},
  \citenamefont {Hook},\ and\ \citenamefont {Balents}}]{burkov2011topological}%
  \BibitemOpen
  \bibfield  {author} {\bibinfo {author} {\bibfnamefont {A.}~\bibnamefont
  {Burkov}}, \bibinfo {author} {\bibfnamefont {M.}~\bibnamefont {Hook}}, \ and\
  \bibinfo {author} {\bibfnamefont {L.}~\bibnamefont {Balents}},\ }\href@noop
  {} {\bibfield  {journal} {\bibinfo  {journal} {Phys. Rev. B}\ }\textbf
  {\bibinfo {volume} {84}},\ \bibinfo {pages} {235126} (\bibinfo {year}
  {2011})}\BibitemShut {NoStop}%
\bibitem [{\citenamefont {Wan}\ \emph {et~al.}(2011)\citenamefont {Wan},
  \citenamefont {Turner}, \citenamefont {Vishwanath},\ and\ \citenamefont
  {Savrasov}}]{wan2011topological}%
  \BibitemOpen
  \bibfield  {author} {\bibinfo {author} {\bibfnamefont {X.}~\bibnamefont
  {Wan}}, \bibinfo {author} {\bibfnamefont {A.~M.}\ \bibnamefont {Turner}},
  \bibinfo {author} {\bibfnamefont {A.}~\bibnamefont {Vishwanath}}, \ and\
  \bibinfo {author} {\bibfnamefont {S.~Y.}\ \bibnamefont {Savrasov}},\
  }\href@noop {} {\bibfield  {journal} {\bibinfo  {journal} {Phys. Rev. B}\
  }\textbf {\bibinfo {volume} {83}},\ \bibinfo {pages} {205101} (\bibinfo
  {year} {2011})}\BibitemShut {NoStop}%
\bibitem [{\citenamefont {Hosur}\ and\ \citenamefont
  {Qi}(2013)}]{hosur2013recent}%
  \BibitemOpen
  \bibfield  {author} {\bibinfo {author} {\bibfnamefont {P.}~\bibnamefont
  {Hosur}}\ and\ \bibinfo {author} {\bibfnamefont {X.}~\bibnamefont {Qi}},\
  }\href@noop {} {\bibfield  {journal} {\bibinfo  {journal} {Comptes Rendus
  Physique}\ }\textbf {\bibinfo {volume} {14}},\ \bibinfo {pages} {857}
  (\bibinfo {year} {2013})}\BibitemShut {NoStop}%
\bibitem [{\citenamefont {Liu}\ \emph {et~al.}(2014{\natexlab{a}})\citenamefont
  {Liu}, \citenamefont {Zhou2}, \citenamefont {Zhang}, \citenamefont {Wang},
  \citenamefont {Weng}, \citenamefont {Prabhakaran}, \citenamefont {Mo},
  \citenamefont {Shen}, \citenamefont {Fang}, \citenamefont {Dai},
  \citenamefont {Hussain},\ and\ \citenamefont {Chen}}]{liu2014discovery}%
  \BibitemOpen
  \bibfield  {author} {\bibinfo {author} {\bibfnamefont {Z.~K.}\ \bibnamefont
  {Liu}}, \bibinfo {author} {\bibfnamefont {B.}~\bibnamefont {Zhou2}}, \bibinfo
  {author} {\bibfnamefont {Y.}~\bibnamefont {Zhang}}, \bibinfo {author}
  {\bibfnamefont {Z.~J.}\ \bibnamefont {Wang}}, \bibinfo {author}
  {\bibfnamefont {H.~M.}\ \bibnamefont {Weng}}, \bibinfo {author}
  {\bibfnamefont {D.}~\bibnamefont {Prabhakaran}}, \bibinfo {author}
  {\bibfnamefont {S.-K.}\ \bibnamefont {Mo}}, \bibinfo {author} {\bibfnamefont
  {Z.~X.}\ \bibnamefont {Shen}}, \bibinfo {author} {\bibfnamefont
  {Z.}~\bibnamefont {Fang}}, \bibinfo {author} {\bibfnamefont {X.}~\bibnamefont
  {Dai}}, \bibinfo {author} {\bibfnamefont {Z.}~\bibnamefont {Hussain}}, \ and\
  \bibinfo {author} {\bibfnamefont {Y.~L.}\ \bibnamefont {Chen}},\ }\href
  {\doibase 10.1126/science.1245085} {\bibfield  {journal} {\bibinfo  {journal}
  {Science}\ }\textbf {\bibinfo {volume} {343}} (\bibinfo {year}
  {2014}{\natexlab{a}}),\ 10.1126/science.1245085}\BibitemShut {NoStop}%
\bibitem [{\citenamefont {Liu}\ \emph {et~al.}(2014{\natexlab{b}})\citenamefont
  {Liu}, \citenamefont {Jiang}, \citenamefont {Zhou}, \citenamefont {Wang},
  \citenamefont {Zhang}, \citenamefont {Weng}, \citenamefont {Prabhakaran},
  \citenamefont {Mo}, \citenamefont {Peng}, \citenamefont {Dudin},
  \citenamefont {Kim}, \citenamefont {Hoesch}, \citenamefont {Fang},
  \citenamefont {Dai}, \citenamefont {Shen}, \citenamefont {Feng},
  \citenamefont {Hussain},\ and\ \citenamefont {Chen}}]{liu2014stable}%
  \BibitemOpen
  \bibfield  {author} {\bibinfo {author} {\bibfnamefont {Z.~K.}\ \bibnamefont
  {Liu}}, \bibinfo {author} {\bibfnamefont {J.}~\bibnamefont {Jiang}}, \bibinfo
  {author} {\bibfnamefont {B.}~\bibnamefont {Zhou}}, \bibinfo {author}
  {\bibfnamefont {Z.~J.}\ \bibnamefont {Wang}}, \bibinfo {author}
  {\bibfnamefont {Y.}~\bibnamefont {Zhang}}, \bibinfo {author} {\bibfnamefont
  {H.~M.}\ \bibnamefont {Weng}}, \bibinfo {author} {\bibfnamefont
  {D.}~\bibnamefont {Prabhakaran}}, \bibinfo {author} {\bibfnamefont {S.-K.}\
  \bibnamefont {Mo}}, \bibinfo {author} {\bibfnamefont {H.}~\bibnamefont
  {Peng}}, \bibinfo {author} {\bibfnamefont {P.}~\bibnamefont {Dudin}},
  \bibinfo {author} {\bibfnamefont {T.}~\bibnamefont {Kim}}, \bibinfo {author}
  {\bibfnamefont {M.}~\bibnamefont {Hoesch}}, \bibinfo {author} {\bibfnamefont
  {Z.}~\bibnamefont {Fang}}, \bibinfo {author} {\bibfnamefont {X.}~\bibnamefont
  {Dai}}, \bibinfo {author} {\bibfnamefont {Z.~X.}\ \bibnamefont {Shen}},
  \bibinfo {author} {\bibfnamefont {D.~L.}\ \bibnamefont {Feng}}, \bibinfo
  {author} {\bibfnamefont {Z.}~\bibnamefont {Hussain}}, \ and\ \bibinfo
  {author} {\bibfnamefont {Y.~L.}\ \bibnamefont {Chen}},\ }\href {\doibase
  10.1038/nmat3990} {\bibfield  {journal} {\bibinfo  {journal} {Nature
  Materials}\ }\textbf {\bibinfo {volume} {13}},\ \bibinfo {pages} {677}
  (\bibinfo {year} {2014}{\natexlab{b}})}\BibitemShut {NoStop}%
\bibitem [{\citenamefont {Brahlek}\ \emph {et~al.}(2012)\citenamefont
  {Brahlek}, \citenamefont {Bansal}, \citenamefont {Koirala}, \citenamefont
  {Xu}, \citenamefont {Neupane}, \citenamefont {Liu}, \citenamefont {Hasan},\
  and\ \citenamefont {Oh}}]{brahlek2012topological}%
  \BibitemOpen
  \bibfield  {author} {\bibinfo {author} {\bibfnamefont {M.}~\bibnamefont
  {Brahlek}}, \bibinfo {author} {\bibfnamefont {N.}~\bibnamefont {Bansal}},
  \bibinfo {author} {\bibfnamefont {N.}~\bibnamefont {Koirala}}, \bibinfo
  {author} {\bibfnamefont {S.-Y.}\ \bibnamefont {Xu}}, \bibinfo {author}
  {\bibfnamefont {M.}~\bibnamefont {Neupane}}, \bibinfo {author} {\bibfnamefont
  {C.}~\bibnamefont {Liu}}, \bibinfo {author} {\bibfnamefont {M.~Z.}\
  \bibnamefont {Hasan}}, \ and\ \bibinfo {author} {\bibfnamefont
  {S.}~\bibnamefont {Oh}},\ }\href@noop {} {\bibfield  {journal} {\bibinfo
  {journal} {Phys. Rev. Lett.}\ }\textbf {\bibinfo {volume} {109}},\ \bibinfo
  {pages} {186403} (\bibinfo {year} {2012})}\BibitemShut {NoStop}%
\bibitem [{\citenamefont {Borisenko}\ \emph {et~al.}(2014)\citenamefont
  {Borisenko}, \citenamefont {Gibson}, \citenamefont {Evtushinsky},
  \citenamefont {Zabolotnyy}, \citenamefont {B{\"u}chner},\ and\ \citenamefont
  {Cava}}]{borisenko2014experimental}%
  \BibitemOpen
  \bibfield  {author} {\bibinfo {author} {\bibfnamefont {S.}~\bibnamefont
  {Borisenko}}, \bibinfo {author} {\bibfnamefont {Q.}~\bibnamefont {Gibson}},
  \bibinfo {author} {\bibfnamefont {D.}~\bibnamefont {Evtushinsky}}, \bibinfo
  {author} {\bibfnamefont {V.}~\bibnamefont {Zabolotnyy}}, \bibinfo {author}
  {\bibfnamefont {B.}~\bibnamefont {B{\"u}chner}}, \ and\ \bibinfo {author}
  {\bibfnamefont {R.~J.}\ \bibnamefont {Cava}},\ }\href@noop {} {\bibfield
  {journal} {\bibinfo  {journal} {Phys. Rev. Lett.}\ }\textbf {\bibinfo
  {volume} {113}},\ \bibinfo {pages} {027603} (\bibinfo {year}
  {2014})}\BibitemShut {NoStop}%
\bibitem [{\citenamefont {Xu}\ \emph {et~al.}(2015{\natexlab{a}})\citenamefont
  {Xu}, \citenamefont {Liu}, \citenamefont {Kushwaha}, \citenamefont {Sankar},
  \citenamefont {Krizan}, \citenamefont {Belopolski}, \citenamefont {Neupane},
  \citenamefont {Bian}, \citenamefont {Alidoust}, \citenamefont {Chang} \emph
  {et~al.}}]{xu2015observation}%
  \BibitemOpen
  \bibfield  {author} {\bibinfo {author} {\bibfnamefont {S.-Y.}\ \bibnamefont
  {Xu}}, \bibinfo {author} {\bibfnamefont {C.}~\bibnamefont {Liu}}, \bibinfo
  {author} {\bibfnamefont {S.~K.}\ \bibnamefont {Kushwaha}}, \bibinfo {author}
  {\bibfnamefont {R.}~\bibnamefont {Sankar}}, \bibinfo {author} {\bibfnamefont
  {J.~W.}\ \bibnamefont {Krizan}}, \bibinfo {author} {\bibfnamefont
  {I.}~\bibnamefont {Belopolski}}, \bibinfo {author} {\bibfnamefont
  {M.}~\bibnamefont {Neupane}}, \bibinfo {author} {\bibfnamefont
  {G.}~\bibnamefont {Bian}}, \bibinfo {author} {\bibfnamefont {N.}~\bibnamefont
  {Alidoust}}, \bibinfo {author} {\bibfnamefont {T.-R.}\ \bibnamefont {Chang}},
   \emph {et~al.},\ }\href@noop {} {\bibfield  {journal} {\bibinfo  {journal}
  {Science}\ }\textbf {\bibinfo {volume} {347}},\ \bibinfo {pages} {294}
  (\bibinfo {year} {2015}{\natexlab{a}})}\BibitemShut {NoStop}%
\bibitem [{\citenamefont {Weng}\ \emph
  {et~al.}(2015{\natexlab{a}})\citenamefont {Weng}, \citenamefont {Fang},
  \citenamefont {Fang}, \citenamefont {Bernevig},\ and\ \citenamefont
  {Dai}}]{weng2015weyl}%
  \BibitemOpen
  \bibfield  {author} {\bibinfo {author} {\bibfnamefont {H.}~\bibnamefont
  {Weng}}, \bibinfo {author} {\bibfnamefont {C.}~\bibnamefont {Fang}}, \bibinfo
  {author} {\bibfnamefont {Z.}~\bibnamefont {Fang}}, \bibinfo {author}
  {\bibfnamefont {B.~A.}\ \bibnamefont {Bernevig}}, \ and\ \bibinfo {author}
  {\bibfnamefont {X.}~\bibnamefont {Dai}},\ }\href@noop {} {\bibfield
  {journal} {\bibinfo  {journal} {Phys. Rev. X}\ }\textbf {\bibinfo {volume}
  {5}},\ \bibinfo {pages} {011029} (\bibinfo {year}
  {2015}{\natexlab{a}})}\BibitemShut {NoStop}%
\bibitem [{\citenamefont {Lv}\ \emph {et~al.}(2015)\citenamefont {Lv},
  \citenamefont {Weng}, \citenamefont {Fu}, \citenamefont {Wang}, \citenamefont
  {Miao}, \citenamefont {Ma}, \citenamefont {Richard}, \citenamefont {Huang},
  \citenamefont {Zhao}, \citenamefont {Chen} \emph
  {et~al.}}]{lv2015experimental}%
  \BibitemOpen
  \bibfield  {author} {\bibinfo {author} {\bibfnamefont {B.}~\bibnamefont
  {Lv}}, \bibinfo {author} {\bibfnamefont {H.}~\bibnamefont {Weng}}, \bibinfo
  {author} {\bibfnamefont {B.}~\bibnamefont {Fu}}, \bibinfo {author}
  {\bibfnamefont {X.}~\bibnamefont {Wang}}, \bibinfo {author} {\bibfnamefont
  {H.}~\bibnamefont {Miao}}, \bibinfo {author} {\bibfnamefont {J.}~\bibnamefont
  {Ma}}, \bibinfo {author} {\bibfnamefont {P.}~\bibnamefont {Richard}},
  \bibinfo {author} {\bibfnamefont {X.}~\bibnamefont {Huang}}, \bibinfo
  {author} {\bibfnamefont {L.}~\bibnamefont {Zhao}}, \bibinfo {author}
  {\bibfnamefont {G.}~\bibnamefont {Chen}},  \emph {et~al.},\ }\href@noop {}
  {\bibfield  {journal} {\bibinfo  {journal} {Phys. Rev. X}\ }\textbf {\bibinfo
  {volume} {5}},\ \bibinfo {pages} {031013} (\bibinfo {year}
  {2015})}\BibitemShut {NoStop}%
\bibitem [{\citenamefont {Huang}\ \emph {et~al.}(2015)\citenamefont {Huang},
  \citenamefont {Xu}, \citenamefont {Belopolski}, \citenamefont {Lee},
  \citenamefont {Chang}, \citenamefont {Wang}, \citenamefont {Alidoust},
  \citenamefont {Bian}, \citenamefont {Neupane}, \citenamefont {Zhang} \emph
  {et~al.}}]{huang2015weyl}%
  \BibitemOpen
  \bibfield  {author} {\bibinfo {author} {\bibfnamefont {S.-M.}\ \bibnamefont
  {Huang}}, \bibinfo {author} {\bibfnamefont {S.-Y.}\ \bibnamefont {Xu}},
  \bibinfo {author} {\bibfnamefont {I.}~\bibnamefont {Belopolski}}, \bibinfo
  {author} {\bibfnamefont {C.-C.}\ \bibnamefont {Lee}}, \bibinfo {author}
  {\bibfnamefont {G.}~\bibnamefont {Chang}}, \bibinfo {author} {\bibfnamefont
  {B.}~\bibnamefont {Wang}}, \bibinfo {author} {\bibfnamefont {N.}~\bibnamefont
  {Alidoust}}, \bibinfo {author} {\bibfnamefont {G.}~\bibnamefont {Bian}},
  \bibinfo {author} {\bibfnamefont {M.}~\bibnamefont {Neupane}}, \bibinfo
  {author} {\bibfnamefont {C.}~\bibnamefont {Zhang}},  \emph {et~al.},\
  }\href@noop {} {\bibfield  {journal} {\bibinfo  {journal} {Nature
  communications}\ }\textbf {\bibinfo {volume} {6}} (\bibinfo {year}
  {2015})}\BibitemShut {NoStop}%
\bibitem [{\citenamefont {Xu}\ \emph {et~al.}(2015{\natexlab{b}})\citenamefont
  {Xu}, \citenamefont {Belopolski}, \citenamefont {Alidoust}, \citenamefont
  {Neupane}, \citenamefont {Bian}, \citenamefont {Zhang}, \citenamefont
  {Sankar}, \citenamefont {Chang}, \citenamefont {Yuan}, \citenamefont {Lee}
  \emph {et~al.}}]{xu2015discovery}%
  \BibitemOpen
  \bibfield  {author} {\bibinfo {author} {\bibfnamefont {S.-Y.}\ \bibnamefont
  {Xu}}, \bibinfo {author} {\bibfnamefont {I.}~\bibnamefont {Belopolski}},
  \bibinfo {author} {\bibfnamefont {N.}~\bibnamefont {Alidoust}}, \bibinfo
  {author} {\bibfnamefont {M.}~\bibnamefont {Neupane}}, \bibinfo {author}
  {\bibfnamefont {G.}~\bibnamefont {Bian}}, \bibinfo {author} {\bibfnamefont
  {C.}~\bibnamefont {Zhang}}, \bibinfo {author} {\bibfnamefont
  {R.}~\bibnamefont {Sankar}}, \bibinfo {author} {\bibfnamefont
  {G.}~\bibnamefont {Chang}}, \bibinfo {author} {\bibfnamefont
  {Z.}~\bibnamefont {Yuan}}, \bibinfo {author} {\bibfnamefont {C.-C.}\
  \bibnamefont {Lee}},  \emph {et~al.},\ }\href@noop {} {\bibfield  {journal}
  {\bibinfo  {journal} {Science}\ }\textbf {\bibinfo {volume} {349}},\ \bibinfo
  {pages} {613} (\bibinfo {year} {2015}{\natexlab{b}})}\BibitemShut {NoStop}%
\bibitem [{\citenamefont {Phillips}\ and\ \citenamefont
  {Aji}(2014)}]{phillips2014tunable}%
  \BibitemOpen
  \bibfield  {author} {\bibinfo {author} {\bibfnamefont {M.}~\bibnamefont
  {Phillips}}\ and\ \bibinfo {author} {\bibfnamefont {V.}~\bibnamefont {Aji}},\
  }\href@noop {} {\bibfield  {journal} {\bibinfo  {journal} {Physical Review
  B}\ }\textbf {\bibinfo {volume} {90}},\ \bibinfo {pages} {115111} (\bibinfo
  {year} {2014})}\BibitemShut {NoStop}%
\bibitem [{\citenamefont {Kim}\ \emph {et~al.}(2015)\citenamefont {Kim},
  \citenamefont {Wieder}, \citenamefont {Kane}, \citenamefont {Rappe} \emph
  {et~al.}}]{kim2015dirac}%
  \BibitemOpen
  \bibfield  {author} {\bibinfo {author} {\bibfnamefont {Y.}~\bibnamefont
  {Kim}}, \bibinfo {author} {\bibfnamefont {B.~J.}\ \bibnamefont {Wieder}},
  \bibinfo {author} {\bibfnamefont {C.}~\bibnamefont {Kane}}, \bibinfo {author}
  {\bibfnamefont {M.}~\bibnamefont {Rappe}},  \emph {et~al.},\ }\href@noop {}
  {\bibfield  {journal} {\bibinfo  {journal} {arXiv preprint arXiv:1504.03807}\
  } (\bibinfo {year} {2015})}\BibitemShut {NoStop}%
\bibitem [{\citenamefont {Yu}\ \emph {et~al.}(2015)\citenamefont {Yu},
  \citenamefont {Weng}, \citenamefont {Fang}, \citenamefont {Dai},\ and\
  \citenamefont {Hu}}]{yu2015topological}%
  \BibitemOpen
  \bibfield  {author} {\bibinfo {author} {\bibfnamefont {R.}~\bibnamefont
  {Yu}}, \bibinfo {author} {\bibfnamefont {H.}~\bibnamefont {Weng}}, \bibinfo
  {author} {\bibfnamefont {Z.}~\bibnamefont {Fang}}, \bibinfo {author}
  {\bibfnamefont {X.}~\bibnamefont {Dai}}, \ and\ \bibinfo {author}
  {\bibfnamefont {X.}~\bibnamefont {Hu}},\ }\href@noop {} {\bibfield  {journal}
  {\bibinfo  {journal} {arXiv preprint arXiv:1504.04577}\ } (\bibinfo {year}
  {2015})}\BibitemShut {NoStop}%
\bibitem [{\citenamefont {Weng}\ \emph
  {et~al.}(2015{\natexlab{b}})\citenamefont {Weng}, \citenamefont {Liang},
  \citenamefont {Xu}, \citenamefont {Yu}, \citenamefont {Fang}, \citenamefont
  {Dai},\ and\ \citenamefont {Kawazoe}}]{weng2015topological}%
  \BibitemOpen
  \bibfield  {author} {\bibinfo {author} {\bibfnamefont {H.}~\bibnamefont
  {Weng}}, \bibinfo {author} {\bibfnamefont {Y.}~\bibnamefont {Liang}},
  \bibinfo {author} {\bibfnamefont {Q.}~\bibnamefont {Xu}}, \bibinfo {author}
  {\bibfnamefont {R.}~\bibnamefont {Yu}}, \bibinfo {author} {\bibfnamefont
  {Z.}~\bibnamefont {Fang}}, \bibinfo {author} {\bibfnamefont {X.}~\bibnamefont
  {Dai}}, \ and\ \bibinfo {author} {\bibfnamefont {Y.}~\bibnamefont
  {Kawazoe}},\ }\href {\doibase 10.1103/PhysRevB.92.045108} {\bibfield
  {journal} {\bibinfo  {journal} {Phys. Rev. B}\ }\textbf {\bibinfo {volume}
  {92}},\ \bibinfo {pages} {045108} (\bibinfo {year}
  {2015}{\natexlab{b}})}\BibitemShut {NoStop}%
\bibitem [{\citenamefont {Ramamurthy}\ and\ \citenamefont
  {Hughes}(2015)}]{ramamurthy2015quasi}%
  \BibitemOpen
  \bibfield  {author} {\bibinfo {author} {\bibfnamefont {S.~T.}\ \bibnamefont
  {Ramamurthy}}\ and\ \bibinfo {author} {\bibfnamefont {T.~L.}\ \bibnamefont
  {Hughes}},\ }\href@noop {} {\bibfield  {journal} {\bibinfo  {journal} {arXiv
  preprint arXiv:1508.01205}\ } (\bibinfo {year} {2015})}\BibitemShut {NoStop}%
\bibitem [{\citenamefont {Aji}(2012)}]{aji2012adler}%
  \BibitemOpen
  \bibfield  {author} {\bibinfo {author} {\bibfnamefont {V.}~\bibnamefont
  {Aji}},\ }\href@noop {} {\bibfield  {journal} {\bibinfo  {journal} {Physical
  Review B}\ }\textbf {\bibinfo {volume} {85}},\ \bibinfo {pages} {241101}
  (\bibinfo {year} {2012})}\BibitemShut {NoStop}%
\bibitem [{\citenamefont {Son}\ and\ \citenamefont
  {Yamamoto}(2012)}]{son2012berry}%
  \BibitemOpen
  \bibfield  {author} {\bibinfo {author} {\bibfnamefont {D.~T.}\ \bibnamefont
  {Son}}\ and\ \bibinfo {author} {\bibfnamefont {N.}~\bibnamefont {Yamamoto}},\
  }\href@noop {} {\bibfield  {journal} {\bibinfo  {journal} {Physical review
  letters}\ }\textbf {\bibinfo {volume} {109}},\ \bibinfo {pages} {181602}
  (\bibinfo {year} {2012})}\BibitemShut {NoStop}%
\bibitem [{\citenamefont {Zyuzin}\ and\ \citenamefont
  {Burkov}(2012)}]{zyuzin2012topological}%
  \BibitemOpen
  \bibfield  {author} {\bibinfo {author} {\bibfnamefont {A.}~\bibnamefont
  {Zyuzin}}\ and\ \bibinfo {author} {\bibfnamefont {A.}~\bibnamefont
  {Burkov}},\ }\href@noop {} {\bibfield  {journal} {\bibinfo  {journal} {Phys.
  Rev. B}\ }\textbf {\bibinfo {volume} {86}},\ \bibinfo {pages} {115133}
  (\bibinfo {year} {2012})}\BibitemShut {NoStop}%
\bibitem [{\citenamefont {Vazifeh}\ and\ \citenamefont
  {Franz}(2013)}]{vazifeh2013electromagnetic}%
  \BibitemOpen
  \bibfield  {author} {\bibinfo {author} {\bibfnamefont {M.~M.}\ \bibnamefont
  {Vazifeh}}\ and\ \bibinfo {author} {\bibfnamefont {M.}~\bibnamefont
  {Franz}},\ }\href {\doibase 10.1103/PhysRevLett.111.027201} {\bibfield
  {journal} {\bibinfo  {journal} {Phys. Rev. Lett.}\ }\textbf {\bibinfo
  {volume} {111}},\ \bibinfo {pages} {027201} (\bibinfo {year}
  {2013})}\BibitemShut {NoStop}%
\bibitem [{\citenamefont {Goswami}\ and\ \citenamefont
  {Tewari}(2013)}]{goswami2013axionic}%
  \BibitemOpen
  \bibfield  {author} {\bibinfo {author} {\bibfnamefont {P.}~\bibnamefont
  {Goswami}}\ and\ \bibinfo {author} {\bibfnamefont {S.}~\bibnamefont
  {Tewari}},\ }\href@noop {} {\bibfield  {journal} {\bibinfo  {journal} {Phys.
  Rev. B}\ }\textbf {\bibinfo {volume} {88}},\ \bibinfo {pages} {245107}
  (\bibinfo {year} {2013})}\BibitemShut {NoStop}%
\bibitem [{\citenamefont {Son}\ and\ \citenamefont
  {Spivak}(2013)}]{son2013chiral}%
  \BibitemOpen
  \bibfield  {author} {\bibinfo {author} {\bibfnamefont {D.}~\bibnamefont
  {Son}}\ and\ \bibinfo {author} {\bibfnamefont {B.}~\bibnamefont {Spivak}},\
  }\href@noop {} {\bibfield  {journal} {\bibinfo  {journal} {Physical Review
  B}\ }\textbf {\bibinfo {volume} {88}},\ \bibinfo {pages} {104412} (\bibinfo
  {year} {2013})}\BibitemShut {NoStop}%
\bibitem [{\citenamefont {Liu}\ \emph {et~al.}(2013)\citenamefont {Liu},
  \citenamefont {Ye},\ and\ \citenamefont {Qi}}]{liu2013chiral}%
  \BibitemOpen
  \bibfield  {author} {\bibinfo {author} {\bibfnamefont {C.-X.}\ \bibnamefont
  {Liu}}, \bibinfo {author} {\bibfnamefont {P.}~\bibnamefont {Ye}}, \ and\
  \bibinfo {author} {\bibfnamefont {X.-L.}\ \bibnamefont {Qi}},\ }\href@noop {}
  {\bibfield  {journal} {\bibinfo  {journal} {Phys. Rev. B}\ }\textbf {\bibinfo
  {volume} {87}},\ \bibinfo {pages} {235306} (\bibinfo {year}
  {2013})}\BibitemShut {NoStop}%
\bibitem [{\citenamefont {Burkov}(2014)}]{burkov2014chiral}%
  \BibitemOpen
  \bibfield  {author} {\bibinfo {author} {\bibfnamefont {A.}~\bibnamefont
  {Burkov}},\ }\href@noop {} {\bibfield  {journal} {\bibinfo  {journal} {Phys.
  Rev. Lett.}\ }\textbf {\bibinfo {volume} {113}},\ \bibinfo {pages} {247203}
  (\bibinfo {year} {2014})}\BibitemShut {NoStop}%
\bibitem [{\citenamefont {McClure}(1956)}]{mcclure1956diamagnetism}%
  \BibitemOpen
  \bibfield  {author} {\bibinfo {author} {\bibfnamefont {J.~W.}\ \bibnamefont
  {McClure}},\ }\href {\doibase 10.1103/PhysRev.104.666} {\bibfield  {journal}
  {\bibinfo  {journal} {Phys. Rev.}\ }\textbf {\bibinfo {volume} {104}},\
  \bibinfo {pages} {666} (\bibinfo {year} {1956})}\BibitemShut {NoStop}%
\bibitem [{\citenamefont {Safran}\ and\ \citenamefont
  {DiSalvo}(1979)}]{safran1979theory}%
  \BibitemOpen
  \bibfield  {author} {\bibinfo {author} {\bibfnamefont {S.}~\bibnamefont
  {Safran}}\ and\ \bibinfo {author} {\bibfnamefont {F.}~\bibnamefont
  {DiSalvo}},\ }\href@noop {} {\bibfield  {journal} {\bibinfo  {journal} {Phys.
  Rev. B}\ }\textbf {\bibinfo {volume} {20}},\ \bibinfo {pages} {4889}
  (\bibinfo {year} {1979})}\BibitemShut {NoStop}%
\bibitem [{\citenamefont {Koshino}\ and\ \citenamefont
  {Ando}(2007{\natexlab{a}})}]{koshino2007diamagnetism}%
  \BibitemOpen
  \bibfield  {author} {\bibinfo {author} {\bibfnamefont {M.}~\bibnamefont
  {Koshino}}\ and\ \bibinfo {author} {\bibfnamefont {T.}~\bibnamefont {Ando}},\
  }\href@noop {} {\bibfield  {journal} {\bibinfo  {journal} {Phys. Rev. B}\
  }\textbf {\bibinfo {volume} {75}},\ \bibinfo {pages} {235333} (\bibinfo
  {year} {2007}{\natexlab{a}})}\BibitemShut {NoStop}%
\bibitem [{\citenamefont {Koshino}\ and\ \citenamefont
  {Ando}(2007{\natexlab{b}})}]{koshino2007orbital}%
  \BibitemOpen
  \bibfield  {author} {\bibinfo {author} {\bibfnamefont {M.}~\bibnamefont
  {Koshino}}\ and\ \bibinfo {author} {\bibfnamefont {T.}~\bibnamefont {Ando}},\
  }\href@noop {} {\bibfield  {journal} {\bibinfo  {journal} {Physical Review
  B}\ }\textbf {\bibinfo {volume} {76}},\ \bibinfo {pages} {085425} (\bibinfo
  {year} {2007}{\natexlab{b}})}\BibitemShut {NoStop}%
\bibitem [{\citenamefont {Koshino}\ and\ \citenamefont
  {Ando}(2010)}]{koshino2010anomalous}%
  \BibitemOpen
  \bibfield  {author} {\bibinfo {author} {\bibfnamefont {M.}~\bibnamefont
  {Koshino}}\ and\ \bibinfo {author} {\bibfnamefont {T.}~\bibnamefont {Ando}},\
  }\href@noop {} {\bibfield  {journal} {\bibinfo  {journal} {Phys. Rev. B}\
  }\textbf {\bibinfo {volume} {81}},\ \bibinfo {pages} {195431} (\bibinfo
  {year} {2010})}\BibitemShut {NoStop}%
\bibitem [{\citenamefont {Ito}\ and\ \citenamefont {Nomura}()}]{ItoNomura}%
  \BibitemOpen
  \bibfield  {author} {\bibinfo {author} {\bibfnamefont {T.}~\bibnamefont
  {Ito}}\ and\ \bibinfo {author} {\bibfnamefont {K.}~\bibnamefont {Nomura}},\
  }\href@noop {} {}\bibinfo {note} {Unpublished}\BibitemShut {NoStop}%
\bibitem [{\citenamefont {Ominato}\ and\ \citenamefont
  {Koshino}(2013)}]{ominato2013orbital}%
  \BibitemOpen
  \bibfield  {author} {\bibinfo {author} {\bibfnamefont {Y.}~\bibnamefont
  {Ominato}}\ and\ \bibinfo {author} {\bibfnamefont {M.}~\bibnamefont
  {Koshino}},\ }\href@noop {} {\bibfield  {journal} {\bibinfo  {journal} {Phys.
  Rev. B}\ }\textbf {\bibinfo {volume} {87}},\ \bibinfo {pages} {115433}
  (\bibinfo {year} {2013})}\BibitemShut {NoStop}%
\bibitem [{\citenamefont {Nomura}\ and\ \citenamefont
  {Nagaosa}(2010)}]{nomura2010electric}%
  \BibitemOpen
  \bibfield  {author} {\bibinfo {author} {\bibfnamefont {K.}~\bibnamefont
  {Nomura}}\ and\ \bibinfo {author} {\bibfnamefont {N.}~\bibnamefont
  {Nagaosa}},\ }\href@noop {} {\bibfield  {journal} {\bibinfo  {journal}
  {Physical Review B}\ }\textbf {\bibinfo {volume} {82}},\ \bibinfo {pages}
  {161401} (\bibinfo {year} {2010})}\BibitemShut {NoStop}%
\bibitem [{\citenamefont {Tserkovnyak}\ \emph {et~al.}(2015)\citenamefont
  {Tserkovnyak}, \citenamefont {Pesin},\ and\ \citenamefont
  {Loss}}]{tserkovnyak2015spin}%
  \BibitemOpen
  \bibfield  {author} {\bibinfo {author} {\bibfnamefont {Y.}~\bibnamefont
  {Tserkovnyak}}, \bibinfo {author} {\bibfnamefont {D.}~\bibnamefont {Pesin}},
  \ and\ \bibinfo {author} {\bibfnamefont {D.}~\bibnamefont {Loss}},\
  }\href@noop {} {\bibfield  {journal} {\bibinfo  {journal} {Phys. Rev. B}\
  }\textbf {\bibinfo {volume} {91}},\ \bibinfo {pages} {041121} (\bibinfo
  {year} {2015})}\BibitemShut {NoStop}%
\end{thebibliography}%
\end{document}